\documentclass[%
 reprint,
superscriptaddress,
nobibnotes,
 amsmath,amssymb,
 aps,
prl,
longbibliography,
citeautoscript,
floatfix,
showkeys
]{revtex4-2}

\usepackage[T1]{fontenc} 
\usepackage[]{graphicx}
\usepackage[utf8]{inputenc}
\usepackage{blindtext}
\usepackage{lmodern}%
\usepackage{hyperref}%
\usepackage{xcolor}%
\usepackage{placeins}
\usepackage{amsmath,amssymb}
\usepackage{bm}
\usepackage{tabularx}
\usepackage{booktabs}
\usepackage{notes2bib}
\usepackage[normalem]{ulem}
\usepackage[normalem]{ulem}
\usepackage{placeins}
\usepackage{amssymb}

\newcommand{\beq}{\begin{equation}\begin{aligned}}
\newcommand{\eeq}{\end{aligned}\end{equation}}

\newcommand{\figref}[2]{\ref{#1}\textsf{#2}}

\newcommand{\dqmp}{Department of Quantum Matter Physics, University of Geneva, 24 Quai Ernest Ansermet, CH-1211 Geneva, Switzerland}
\newcommand{\gap}{Department of Applied Physics, University of Geneva, 24 Quai Ernest Ansermet, CH-1211 Geneva, Switzerland}

\newcommand{\modena}{Dipartimento di Scienze Fisiche, Informatiche e Matematiche, University of Modena and Reggio Emilia, IT-41125 Modena, Italy}
\newcommand{\cnr}{Centro S3, CNR Istituto Nanoscienze, IT-41125 Modena, Italy}

\definecolor{linkcol}{rgb}{0,0,0.4}
\definecolor{citecol}{rgb}{0.5,0,0}
\definecolor{harvardcrimson}{rgb}{0.79, 0.0, 0.09}
\definecolor{lava}{rgb}{0.81, 0.06, 0.13}

\hypersetup{
	bookmarksopen=true,
	pdftitle="Influence of magnetism on vertical hopping transport in CrSBr",
	pdfauthor="Xiaohanwen Lin et al",
	pdfsubject="", 
	pdfstartview={FitH}, 
	pdfhighlight=/O, 
	colorlinks=true, 
	pdfpagemode=UseNone, 
	pdfpagelayout=SinglePage, 
	pdffitwindow=true, 
	linkcolor=harvardcrimson, 
	citecolor=harvardcrimson, 
	urlcolor=lava
}

	\begin{abstract} 
We investigate the $c$-direction conduction in CrSBr in the linear regime, not accessible in other van der Waals (vdW) magnetic semiconductors, because of the unmeasurably low current. The resistivity --$10^8$-to-$10^{11}$ times larger than in the $a$- and $b$-directions-- exhibits magnetic state dependent thermally activated and variable range hopping transport. In the spin-flip phase at 2~T, the activation energy is 20~meV lower than in the antiferromagnetic state, due to a downshift of the conduction band edge, in agreement with ab-initio calculations. In the variable range hopping regime, the average hopping length decreases from twice the interlayer distance to the interlayer distance at 2~T, because in the antiferromagnetic state the large exchange energy impedes electrons hopping between adjacent layers. Our work demonstrates that the linear transport regime provides new information about electronic processes in vdW magnetic semiconductors, and shows how magnetism influences these processes both in real and reciprocal space. 
 \end{abstract}

\begin{document}
	

	\title{Influence of magnetism on vertical hopping transport in CrSBr}
	\author{Xiaohanwen Lin} 
 \affiliation{\dqmp}
 \affiliation{\gap}
 \author{Fan Wu} 
 \affiliation{\dqmp}
 \affiliation{\gap}

 \author{Sara A. Lopéz‐Paz}
 \affiliation{\dqmp}
  \author{Fabian O. von Rohr} 
 \affiliation{\dqmp}


	\author{Marco Gibertini} 
 \affiliation{\modena}
 \affiliation{\cnr}
 \author{Ignacio Gutiérrez-Lezama} 
 \author{Alberto F. Morpurgo}
 \email{alberto.morpurgo@unige.ch}
 \affiliation{\dqmp}
 \affiliation{\gap}
 
 	\date{\today}
 \maketitle
\section*{Introduction}
Temperature and magnetic field-dependent transport measurements provide a powerful approach to mapping the magnetic phase diagram of two-dimensional magnetic semiconductors (2DMSs)~\cite{burch_magnetism_2018,gibertini_magnetic_2019,mak_probing_2019,wang_determining_2019,long_persistence_2020}. In most cases, measurements rely on devices with electrodes attached on opposite sides of a multilayer, to probe the current flowing perpendicular to the layers~\cite{gong_discovery_2017,huang_layer-dependent_2017,klein_probing_2018,song_giant_2018,kim_one_2018,wang_electric-field_2018,kim_tailored_2019,song_switching_2019,wang_determining_2019,long_persistence_2020,wang_magnetization_2021,soler-delgado_probing_2022}. The resulting current-voltage ($I$-$V$) curves commonly exhibit a characteristic strong non-linearity, with $ln(I/V^2)$ scaling proportionally to $1/V$ at high bias, a dependence usually interpreted in terms of a Fowler-Nordheim (FN) tunneling scenario (see Fig.~\figref{fig:1} (a))~\cite{fowler1928electron,lenzlinger1969fowler}. In its usual form, however, FN tunneling theory assumes electrons to have a quadratic dispersion relation up to energies much higher than the barrier height, an assumption invalid for most 2DMSs explored so far, because of their extremely narrow bandwidths~\cite{wang_electronic_2011,zheng_ab_2019}. Interpreting experiments in terms of FN tunneling should be therefore considered as a useful phenomenological approach, with the microscopic electronic processes responsible for vertical transport in 2D magnetic semiconductors that remain to be identified.~\\

Gaining a deeper understanding of these microscopic processes experimentally is difficult because most multilayers of 2DMSs are so resistive that the current in the linear $I$-$V$ transport regime –-simpler to interpret than the non-linear one–- is too small to be detected~\cite{wang_electric-field_2018,wang_determining_2019,long_persistence_2020,soler-delgado_probing_2022}. Even though for truly atomically thin layers linear transport due to direct tunneling is observed, current in the linear transport regime becomes unmeasurably small for barriers that are only a few monolayers thick. To probe more effectively the microscopic transport processes in 2DMSs and their interplay with the magnetic state, materials are needed in which linear transport can be measured in a broad range of temperatures and magnetic fields, to test whether the evolution of the conductivity can be properly described in terms of processes well-understood in conventional semiconductors.~\\

Here we report on vertical transport experiments on CrSBr exfoliated layers of different thickness, in which the linear regime is detected over a broad range of temperatures and magnetic fields. We find that the linear resistance increases in a thermally activated way ($R=R_{0} \exp \left(\frac{E_{a}}{kT}\right)$; see Fig.~\figref{fig:1} (a)) between room temperature and approximately 60~K, before crossing over to a regime dominated by variable range hopping (VRH, $R=R_{0} \exp \left(\frac{T_{0}}{T}\right)^{0.5}$; see Fig.~\figref{fig:1}(a)) between 40~K and 10~K, both for $ \mu_0H = 0$ T --in the antiferromagnetic state of the material-- and at $ \mu_0H = 2$ T --in the spin-flip phase with all spins aligned. The activation energy $E_\mathrm{a}$ and characteristic hopping temperature $T_\mathrm{0}$ decrease when a magnetic field is applied. We perform an analysis in terms of established theory for hopping transport and attribute the change in $E_\mathrm{a}$ to a 20~meV downshift of the conduction band edge, and the reduction in $T_\mathrm{0}$ to the possibility for the electrons to do shorter hops in the spin-aligned state. These conclusions show how the analysis of linear transport provides detailed information about the interplay of the magnetic state and microscopic electronic processes in van der Waals(vdW) magnetic semiconductors.

\section*{RESULTS and ANALYSIS}
\begin{figure}[t]
 \includegraphics[width=\linewidth]{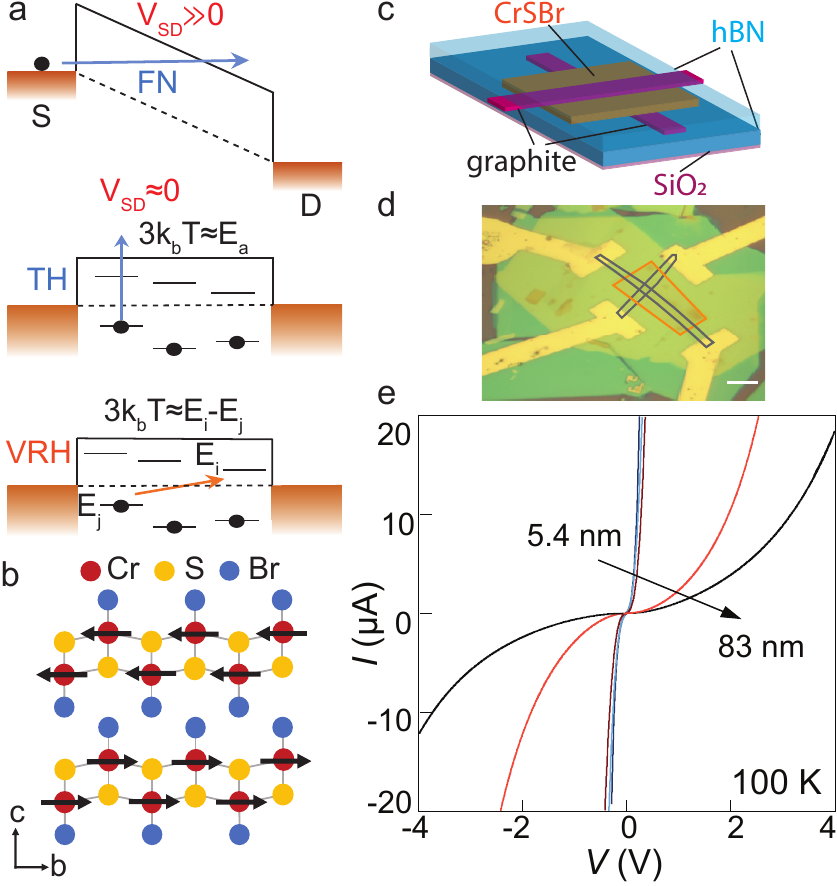}
 \caption{(a) Schematic representation of possible vertical transport mechanisms. Band diagram of Fowler-Nordheim (FN) tunneling of electrons through a triangular potential barrier at high applied source-drain bias $V_{sd}$ . At low applied bias --in the linear regime-- electron transport occurs either via thermally activated hopping in the high-temperature regime (middle panel), or via variable range hopping (VRH) at lower temperatures (bottom panel). (b) Crystal structure of bi-layer CrSBr viewed along the a axis. The red, blue, and yellow balls represent the Cr, Br and S atoms, respectively. The black arrows represent the individual spins on the Cr atoms, showing the layered antiferromagnetic alignment within the material. (c) Schematic representation and (d) optical microscope image of one of the CrSBr devices employed in our experiments. The device consists of a CrSBr crystal (orange contour in (d)) placed in-between graphite contacts (gray contour in (d)) encapsulated with top and bottom hexagonal-boron nitrite (hBN). The scale bar in (d) is 5 $\mu$m. (e) $I-V$ characteristics of five different CrSBr devices, with thickness 5.4, 5.4, 6.3, 53, 83~nm, measured at 100 K.}
 \label{fig:1}
\end{figure}

CrSBr is a vdW layered antiferromagnetic semiconductor (see Fig.~\figref{fig:1}(b)) with relatively large bandwidth within the layers~\cite{wilson_interlayer_2021,wu_quasi-1d_2022}, and  Néel temperature of $T_N=$ 132 K~\cite{telford_layered_2020,wilson_interlayer_2021,wu_quasi-1d_2022,boix-constant_probing_2022}. Its magnetic properties have been extensively characterized both for bulk and for thin exfoliated layers, and it is established that, at low temperature, the application of a magnetic field of 2 T parallel to the $c$-axis brings the material in the fully spin-polarized state~\cite{wu_quasi-1d_2022,wilson_interlayer_2021,telford_layered_2020}. The in-plane transport properties have also been carefully studied experimentally in bulk~\cite{telford_layered_2020} and in field-effect transistors based on exfoliated crystals ~\cite{wu_quasi-1d_2022,telford_coupling_2022}. The conductivity of the material was found to remain high down to cryogenic temperatures, indicating that a large concentration of unintentional dopants (associated to defects in the material) is present, and that the Fermi level is located very close to the conduction band edge. Although important aspects of the experimental results remain to be understood --such as the anomalous, gate-dependent anisotropy in the conductivity-- the temperature dependence of the electrical conductivity in bulk crystals is successfully reproduced by established theory for hopping transport, with thermally activated and variable range hopping regime at high and low temperature~\cite{wu_quasi-1d_2022,telford_coupling_2022}. As for transport in the $c$-direction (i.e., in the direction perpendicular to the layers), only experiments on mono and bilayers have been reported, in which  current is due to direct tunnelling~\cite{boix-constant_probing_2022}. \\

Here we investigate the processes mediating transport through crystals with thickness ranging from approximately 5 to 100 nm, using vertical junction devices in which an exfoliated layer is contacted on opposite sides with few-layer graphene strips (see Fig.~\figref{fig:1} (c) and (d)). The high unintentional doping level of CrSBr crystals is ideal to generate a sufficiently high electrical hopping conductivity also along the $c$-axis, enabling the linear transport regime to be accessed down to low temperature. Differently from the in-plane electrical conductivity,  transport in the $c$-direction is mediated by electrons  hopping between  states in different layers that --depending on the magnetic state of the material-- may have opposite magnetization. To properly model these interlayer hopping processes is important to understand the nature of the defect states involved, and the relation between their spin and the layer magnetization. This information has been obtained in recently reported scanning tunneling microscopy and spectroscopy measurements~\cite{Kleinacsnano_2023}, which revealed the presence of a large density of defects --mainly Br vacancies-- that create states localized within individual layers, with energy close to the conduction band edge. Theoretical calculations within the same work~\cite{Kleinacsnano_2023} also find that carriers in these localized states have their spin aligned with that of the layer magnetization, a conclusion consistent with the interpretation of several experiments reported in the literature~\cite{Sara_2022,telford_coupling_2022,boix-constant_probing_2022}. We will refer back to the nature of the defect-induced states involved in the hopping process later, when we interpret microscopically the results of our measurements.\\

The CrSBr crystals employed in our work were grown by a chemical vapor transport method utilizing S$_2$Br$_2$ as a transport agent~\cite{wu_quasi-1d_2022} and the devices are realized employing by now common pick-up and transfer techniques~\cite{deng_two-dimensional_2021}, electron-beam lithography, electron-beam evaporation of a Pt/Au, and lift-off (for detail of the structure fabrication see Appendix A). We performed measurements on five different devices based on four different exfoliated crystals with different thickness, between room-temperature and 10~K (the precise range depends on multilayer thickness), and found a fully consistent temperature and magnetic field dependent behavior of the resistivity, which we illustrate with data measured on a device in which the CrSBr layer is 5.4 nm thick (data from other devices are shown in Appendix A to C). The consistency of measurements performed on devices of different thickness is important not only to establish the reproducibility of the resistivity data, but also to ensure that contact effects --which would give a more sizable contribution to the resistance in thinner devices-- are not influencing the final results.\\

\begin{figure}[t!]
 \includegraphics[width=\linewidth]{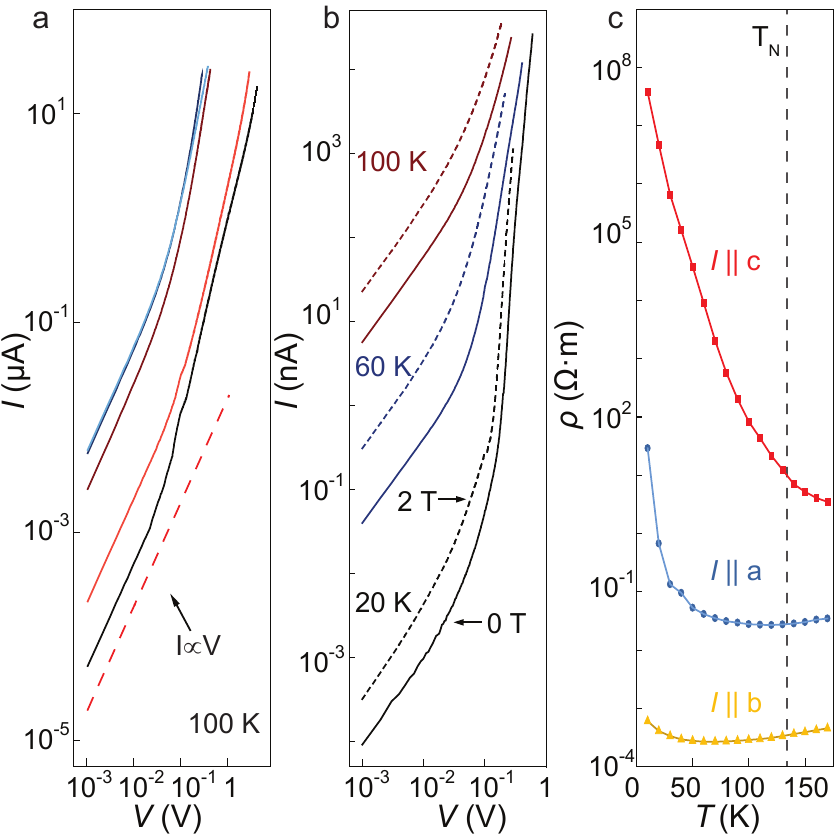}
 \caption{(a) $I-V$ characteristics measured at 100 K on five different CrSBr devices, with thickness 5.4, 5.4, 6.3, 53, 83~nm, plotted in double-logarithmic scale (same data as in Fig.~\figref{fig:1}(e) plotted for V > 0~V). The red dashed line has slope equal to one as expected for a linear $I-V$ dependence. (b) $I-V$ characteristics of a 5.4~nm thick device measured at 20, 60 and 100~K; the solid and dashed lines correspond to measurements performed at $\mu_0H = 0~T$ (in the antiferromagnetic phase of CrSBr) and at $\mu_0H = 2~T$ (in the fully spin-polarized phase). (c) Temperature dependence of the linear resistivity of CrSBr measured along the $a$-, $b$- and $c$-direction (the resistivity in the $a$~and $b$~direction is reproduced from our earlier study of CrSBr field-effect transistors~\cite{wu_quasi-1d_2022}, where we show that the $a$ ($b$) direction corresponds to the long (short) lateral side of the exfoliated crystals.)} 
 \label{fig:2}
\end{figure}

Fig.~\figref{fig:1}(e) shows the $I-V$ characteristics of the five devices studied, measured at $T=100$ K, all exhibiting a clear non-linearity as the bias is increased. We identify the linear regime by plotting the $I-V$ characteristics in a double-logarithmic scale (Fig. ~\figref{fig:2}(a)) and selecting the low-bias part of the curve with slope equals to 1. As expected, in all devices current in this regime rapidly decreases (and the onset of the non-linearity shifts to lower applied voltage) as temperature is lowered. Also, in all devices the application of a magnetic field of 2~T --sufficient to enter the spin-flip phase of CrSBr-- leads to a current increase at all temperatures and biases (see Fig.~\figref{fig:2}(b)). ~\\

\begin{figure}[t!]
 \includegraphics[width=\linewidth]{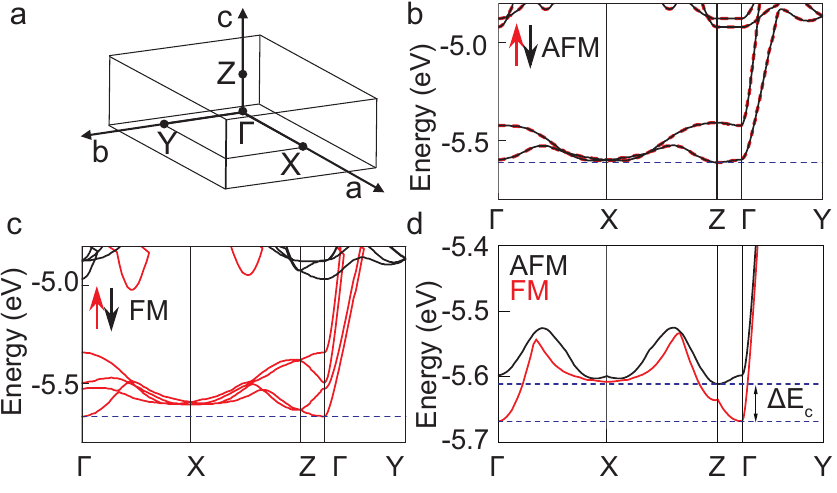}
 \caption{(a) Brillouin zone and high symmetry paths along which the bands are shown in the band structure calculations of antiferromagnetic (b) and ferromagnetic state (c) in CrSBr bulk. The crystallographic $a$, $b$ and $c$-axis correspond to the $\Gamma$-to-$X$, $\Gamma$-to-$Y$ and $\Gamma$-to-$Z$ directions, respectively. The bandwidth along $\Gamma$-to-$Z$ is clearly smaller than in the other two directions. The horizontal dark blue dashed lines in (b) and (c) represent the bottom of the conduction band. (d) Comparison of the conduction band edge position in the antiferromagnetic and ferromagnetic state. A downshift of the conduction band edge ($\bigtriangleup E_{C} \approx 50~meV$) occurs when going from the antiferromagnetic to the ferromagnetic state.}
 \label{fig:4}
\end{figure}

We start our analysis by looking at the magnitude of the linear resistivity, and estimate the electronic anisotropy by comparing the value of the resistivity in the $c-$direction to that in the $a$- and $b$-directions, measured earlier using field-effect transistor devices (we take the value of the resistivity measured at zero gate voltage)~\cite{wu_quasi-1d_2022}. At low temperature, the resistivity in the $c$-direction is approximately 11 orders of magnitude larger than in the $b$-direction, and 6-to-8 orders of magnitude larger than the conductivity in the $a$-direction (the exact value depends on the device from which the values of the $a-$ and $b-$resistivity are extracted; see Fig.~\figref{fig:2}(c)). This giant anisotropy is consistent with the bandwidth in the three different directions, as obtained from ab-initio calculations (see Fig.~\figref{fig:4}(a) and (b)). Specifically, in the antiferromagnetic state, the dispersion in the $\Gamma$-to-$Z$ direction (i.e., the $c$ direction) shows a bandwidth of only a few meV, much smaller than the bandwidth in the other two directions ($\Gamma$-to-$X $ and $\Gamma$-to-$Y$), corresponding respectively to the $a$ and $b$ directions, i.e., to in-plane motion within the material layers. The anisotropy inferred from the resistivity measured in the $c$-direction is extremely large, much larger than in any other van der Waals material that we know of~\cite{najmaei_cross-plane_2018}, and implies a very small electronic coupling between adjacent layers of CrSBr. \\
We then analyze the evolution of the resistance as a function of $T$ in terms of established theory for hopping transport in disordered semiconductors. When the Fermi level is in the tail of states below the conduction band edge, theory predicts the resistance to exhibit a thermally activated behavior $R \propto \exp \left(\frac{E_{a}}{kT}\right)$ at sufficiently high temperature (see Fig.~\figref{fig:1}(a))\cite{shklovskii2013electronic}. The value of $E_a$ may determined by the energy distribution of localized states in the band tail or by the distance between the Fermi level $E_F$ and the conduction band edge (as electrons may have enough energy to hop from a localized state into the conduction band and propagate before being trapped again). At lower temperatures, transport should cross over to a VRH regime, in which $R \propto \exp \left(\frac{T_{0}}{T}\right)^{p}$ (see Fig.~\figref{fig:1}(a)), with $p$ determined by electronic dimensionality, influence of Coulomb repulsion, and other details \cite{shklovskii2013electronic, pollak1991hopping}.\\

\begin{figure}
\centering
 \includegraphics[width=\linewidth]{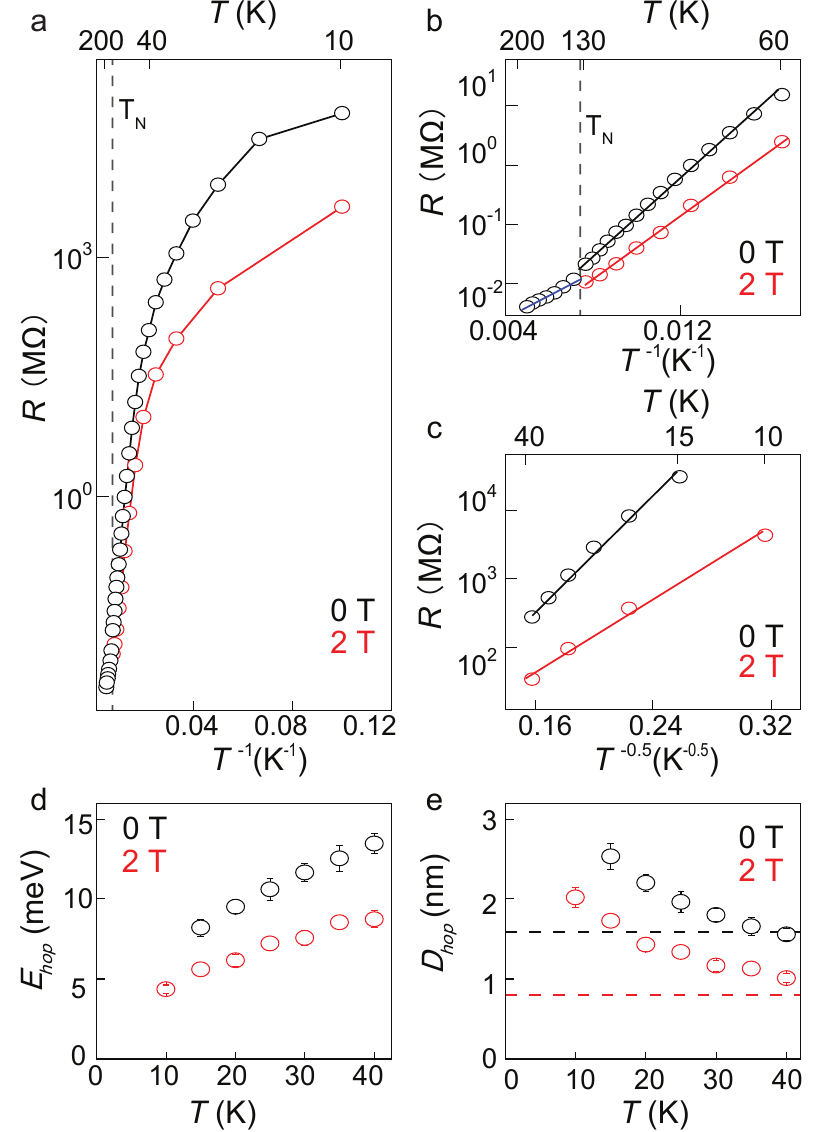}
 \caption{ (a) Resistance of a 5.4~nm thick CrSBr device as a function of $1/T$, measured at $\mu_0H =$ 0~T (black symbols) and 2~T (red symbols; $H\parallel c$) in the full temperature range, covering the thermally activated and variable range hopping regimes. (b) Resistance as a function of $1/T$, focusing on the thermally activated regime between 60 to 200~K. The different symbols correspond to the paramagnetic (blue), antiferromagnetic (black) and spin-flip (red) states of the material (the continuous lines of the corresponding color are linear fits to the data). (c) Resistance as a function of ${T}^{-0.5}$, in the variable range hopping regime between 40 K and 10 K in the antiferromagnetic (black symbols) and in the spin-flip (red symbols) states at $\mu_0H =$ 0 and 2~T, respectively. The continuous lines are linear fits to the data. The analysis of transport in the variable range hopping regime in the temperature range between 10 K and 40 K gives --through Eqs. (1) and (2) in the main text-- the mean hopping energy $E_{hop}$ (d) and the mean hopping distance $D_{hop}$ (e). Black and red symbols correspond to the average values of these quantities (from all devices) extracted from the antiferromagnetic state of CrSBr at $\mu_0H$ = 0~T and in the fully spin-polarized spin-flip phase at $\mu_0H$ = 2~T (see Appendix C for data from different devices). In (e), the dashed horizontal lines represent the distance between two adjacent layers (red dashed line) and to the next-next layer (black dashed line), which correspond to the shortest possible hopping distance in the fully polarized spin-flip state and in the antiferromagnetic state of CrSBr.} 
 \label{fig:3}
\end{figure}

Fig.~\figref{fig:3}(a) shows the evolution of the resistance upon cooling, starting from 200~K (other devices were measured starting from 290~K, see Appendix C). As expected, between 200~K and 60~K $ln(R)$ depends linearly on $1/T$ (see Fig.~\figref{fig:3}(b)), with a change in slope for $T>T_N$ (in the paramagnetic state) and $T<T_N$ (in the layered antiferromagnetic state; see blue and black lines in Fig.~\figref{fig:3}(b)). Additionally, for $T<T_N$, the resistance continues to exhibit a thermally activated behavior also in the presence of a perpendicular magnetic field $\mu_0 H=2$~T --larger than the spin-flip field for all temperatures-- with an activation energy in the spin-flip phase that is smaller than in the antiferromagnetic state (see red line in Fig.~\figref{fig:3}(b)). Linear fits to the data allow the activation energy to be extracted in all different conditions. For this device, we find $E_a=42$ meV for $T$ between 200 K and $T_N$ (i.e., in the paramagnetic state), and $E_a=$ 74~meV and 56~meV for $T$ between $T_N$ and 60~K, respectively without and with a 2~T applied magnetic field. These values are comparable to those found in other devices (see Table \ref{tab1} and Appendix A for details about the different devices), from which we estimate average $E_a = 46\pm 9$ meV in the paramagnetic state(PM), $E_a = 79\pm 6$ meV in the antiferromagnetic state(AFM), and $E_a = 59\pm 4$ meV in the spin-flip phase at $\mu_0H = 2$~T(i.e., ferromagnetic state(FM)).\\
\begin{table}[t]
  \centering
    \caption{Overview of hopping transport parameters: activation energy and $T_0$ value for all measured devices, in the different magnetic states.}
   \begin{tabular}{p{70pt}p{20pt}p{20pt}p{20pt}p{20pt}p{20pt}p{60pt}}
    \toprule
    \toprule
     Device&  A1& A2 &  B &  C & D & Average value\\
    Thickness (nm) & 5.4  & 5.4 & 6.3 & 53 & 83 & / \\
     $E_{a}$-AFM (meV) & 74 & 71 & 80 & 85 & 84 & 79 \\
     $E_{a}$-FM (meV) & 56 & 57 & 58 & 67 & 58 & 59\\  
    $E_{a}$-PM (meV) & 42 & 35 & 54 & 56 & 43 & 46\\
    $T_{0}$-AFM (K) & 2209 & 2663 & /  & 2490 & / & 2454\\ 
    $T_{0}$-FM (K) & 894 & 1122 & /  & 1075 & / & 1030\\
    \bottomrule
    \bottomrule 
    \end{tabular}%
  \label{tab1}%
\end{table}%

The quantitative analysis of the resistance in the VRH regime, with $R \propto \exp \left(T_{0}/{T}\right)^{p}$, depends on the exponent $p$ in the exponential. For isotropic materials theory predicts $p=1/(D+1)$ (where $D$ is the system dimensionality) if electron-electron interactions can be neglected (Mott regime), and $p=1/2$ irrespective of $D$, if electron-electron interaction dominate (Efros-Skhlovskii --ES-- regime). Here, we select $p=1/2$, because the range of temperatures in which the VRH regime is observed --well below 100 K-- corresponds to characteristic energies smaller than 10 meV, and on this energy scale it should certainly be expected that electron-electron interactions play a role in a layered van der Waals semiconductor. Fig.~\figref{fig:3}(c) shows that the relation between $\ln(R)$ and $1/T^{1/2}$ is indeed linear between 40 and 10 K, both in the antiferromagnetic state (black line) and in the spin-flip phase with 2 T applied magnetic field (red line). The values of $T_0$ are extracted by determining the slope, from which we obtain $T_0(B = 0~T)=2209~K$ and $T_0(B=2~T)=894~K$ (see Table \ref{tab1} for the fitting results for the other devices). \\

As the values of $T_0$ give little physical intuition, we resort to VRH theory to relate $T_0$ to quantities that have a direct physical interpretation. These are the average change in energy of an electron upon hopping $E_{hop}$, and the mean distance in a hopping process $D_{hop}$. Within ES theory,\cite{shklovskii2013electronic, pollak1991hopping} these quantities read:
\begin{equation}
E_{h o p}=\frac{1}{2} k T^\frac{1}{2} T_{0}^\frac{1}{2},
\label{Ehop}
 \end{equation}
 \begin{equation}
D_{h o p}=\frac{1}{4} a\left(\frac{T_{0}}{T} \right)^\frac{1}{2}.
\label{Ehop2}
 \end{equation}
Here, $k$ is the Boltzmann constant and $a$ is the localization length (i.e., the spatial extension of the localized states). To proceed with a quantitative analysis, we need to establish the value of $a$. To this end we recall that the localized states involved in the hopping processes originate from defects (mainly Bromine vacancies) in individual layers, whose wave functions are also localized within individual layers (i.e., the overlap of states in neighbouring layers is small). The very high resistivity  in the $c$-direction observed in the experiments is a direct consequence of the small overlap of wavefunctions in different layers, the same reason why the  width of the CrSBr bands in the $c$-direction is extremely narrow.  As the defect states mediating the hopping processes  are localized within the individual layers and only very weakly coupled, we can take as value for the localization length $a$ the distance between two adjacent layers (i.e., $a=0.8$ nm) \\

The values of $E_{hop}$ and $D_{hop}$ extracted from different devices are very close to each other irrespective of thickness of the CrSBr layer. Fig.~\figref{fig:3}(d) and (e) show the temperature dependence of these quantities plotted from Eqs. (1) and (2), using for $T_0$ the average of the values obtained experimentally from  all devices measured. $E_{hop}$ scales approximately linearly with $T$ throughout the range in which VRH properly describes transport. $E_{hop} \approx 4-9$~meV in the spin polarized state, and somewhat larger in the antiferromagnetic state. Finding a value just slightly larger than the thermal energy (i.e., $\simeq 3.5 k_BT$) should be expected and confirms the soundness of the analysis. The hopping distance $D_{hop}$ at the onset of VRH ($T = 40$~K) is close to 1 nm (comparable to the interlayer separation 0.8 nm as illustrated by the red dashed line) in the spin polarized state and is twice the value in the antiferromagnetic state (black dashed line). For both states, $D_{hop}$ increases slightly upon cooling from 40 to 15~K as electrons have to jump further to find states with a appropriate energy (see also discussion below). \\

Before discussing the microscopic interpretation of these findings, we note that these results also indicate the consistency of the $p=0.5$ value selected for the VRH analysis. Indeed, we analyzed the data with smaller $p$ values and found larger hopping distances. For instance for $p=0.25$, $D_{hop}$ comes out to be larger than the thickness of our thinnest device (5.4~nm), a conclusion not internally consistent, since it would imply that electrons can tunnel directly through the entire multilayer, and transport would not be in the hopping regime (and should therefore not show hopping temperature dependence).\\

The 20~meV decrease in activation energy $E_a$ observed when passing from the antiferromagnetic to the spin polarized state of CrSBr, as well as the change in $T_0$ (with the corresponding change in $D_{hop}$ inferred from the analysis of hopping transport), result from the interplay between the magnetic state of CrSBr and electronic transport. As for $E_a$, ab-initio calculations (see Fig.~\figref{fig:4}) do indeed predict that in the ferromagnetic state the conduction band edge is downshifted by a few tens of meV (see Fig.~\figref{fig:4}(d); the exact value depends on the thickness of the multilayer considered). The lower energy of the conduction band edge explains the lowering of the activation energy measured in the presence of a 2~T applied field (which brings CrSBr in the spin-flip phase with all spin aligned), and captures the correct order of magnitude of the effect. Note that a comparable magnitude downshift of the conduction band edge between the antiferromagnetic and the spin-flip phase has also been recently found in CrPS$_4$ --another 2D layered antiferromagnetic semiconductor-- both theoretically and experimentally~\cite{wu_gate-controlled_2023,wu_magnetism-induced_2023}. Our findings on CrSBr therefore suggest that a downshift of the conduction band is a common property of these layered antiferromagnetic materials.\\

The reason why $T_0$ changes when passing from the antiferromagnetic to the spin-flip phase can be easily understood if we consider the corresponding change in average hopping distance. As discussed above, the defects states involving in the hopping process have their spin aligned with the magnetization of the layer hosting them. In the antiferromagnetic state, therefore, electrons in one layer cannot hop to the nearest adjacent layers, because the large exchange energy (approximately 0.6~eV according to ab-initio calculations; see Fig.~\figref{fig:4}(c)) shifts states with the same spin to much larger energy values, and make them energetically inaccessible. As a result,  the shortest hop possible in the $c$-direction in the antiferromagnetic state is to the next-next layer. In contrast, in the spin polarized state hopping to the next layer is unimpeded. That is why the typical hopping distance at $\mu_0H= 2$ T (i.e., in the spin-flip phase) corresponds to the thickness of a single monolayer, and at $\mu_0H = 0$ T in the antiferromagnetic state is twice as long. This is indeed what is quite precisely observed at the largest temperature of the VRH range hopping regime, when electrons have enough energy to find the closest possible state to hop to. As $T$ is lowered the hopping distance increases --while remaining of the same order of magnitude-- because electrons need to do hops a somewhat longer distance, as the hopping sites with suitable energy in nearby layers are displaced laterally relatively to each other (i.e., in general, defects in nearby layers are shifted relatively to each other).\\

\section*{ CONCLUSIONS}
We conclude that the analysis of hopping transport in the direction perpendicular to the layers in CrSBr barriers provides a detailed insight about the nature of the microscopic processes responsible for electronic motion in a layered antiferromagnetic semiconductor. In particular, the evolution with magnetic field of the thermally activated and of the variable range hopping regimes reveals how layered antiferromagnetism influences both the conduction band in reciprocal space and electronic motion in real space. It is certainly worth emphasizing that it is the possibility to analyze transport in terms of an established theory --the theory of hopping transport through disordered semiconductors-- that allows extracting precise and quantiatively reliable information about the the interplay of electron dynamics and the magnetic state of the material.~\\ 

\section*{Acknowledgements}
The authors gratefully acknowledge Alexandre Ferreira for continuous and valuable technical support. We thank Menghan Liao and Nicolas Ubrig for fruitful discussions. AFM Acknowledges financial support from the Swiss National Science Foundation under project 200020\_178891 and the EU Graphene Flagship project for support. MG acknowledges support from the Italian Ministry for University and Research through the PNRR project ECS\_00000033\_ECOSISTER.
FvR acknowledges financial support by the Swiss National Science Foundation under Grant PCEPF2\_1941183.\\

\section{APPENDIX A: Methods} \label{Methods}

\subsection{First-principles calculations}
In Fig.~\figref{fig:4}(b) to (d) of the main text we show the calculated band structure of CrSBr in the layered antiferromagnetic state and in the fully spin-aligned (ferromagnetic) state. The band structure is obtained by first-principles density-functional theory calculations using the Quantum ESPRESSO  distribution~\cite{giannozzi_quantum_2009,giannozzi_advanced_2017}. Van der Waals interactions have been included by adopting the spin-polarised extension\cite{thonhauser_spin_2015} of the revised vdw-DF3-opt2 exchange-correlation functional~\cite{chakraborty_next_2020}, with pseudopotentials from the Standard Solid-State Pseudopotential (SSSP) accuracy library (v1.0)~\cite{prandini_precision_2018,garrity_pseudopotentials_2014} with energy cutoffs of 40~Ry and 320~Ry for wave functions and density, respectively. 
A uniform $12\times9\times4$ mesh of k-points corresponding to a $\Gamma$-centered Monkhorst-Pack grid has been used in self-consistent calculations. Atomic positions and lattice parameters have been fully relaxed within the antiferromagnetic configuration (with doubled unit cell in the vertical direction) using a BFGS algorithm until the residual force on each atom was below 2.6~meV/\AA\ and each component of stress below 0.5 kbar, which gives  lattice parameters in good agreement with experiments ($a=3.52$~\AA, $b=4.71$~\AA, $c=15.96$~\AA). For better comparison with the antiferromagnetic case, the same doubled unit cell has been adopted also for band structure calculations in the ferromagnetic configuration. Energy bands in the two magnetic configurations have been aligned so that the energies of the deepest semicore states match.
\subsection{Device Fabrication}

 \begin{figure*}[htbp]
  \includegraphics[width=1\linewidth]{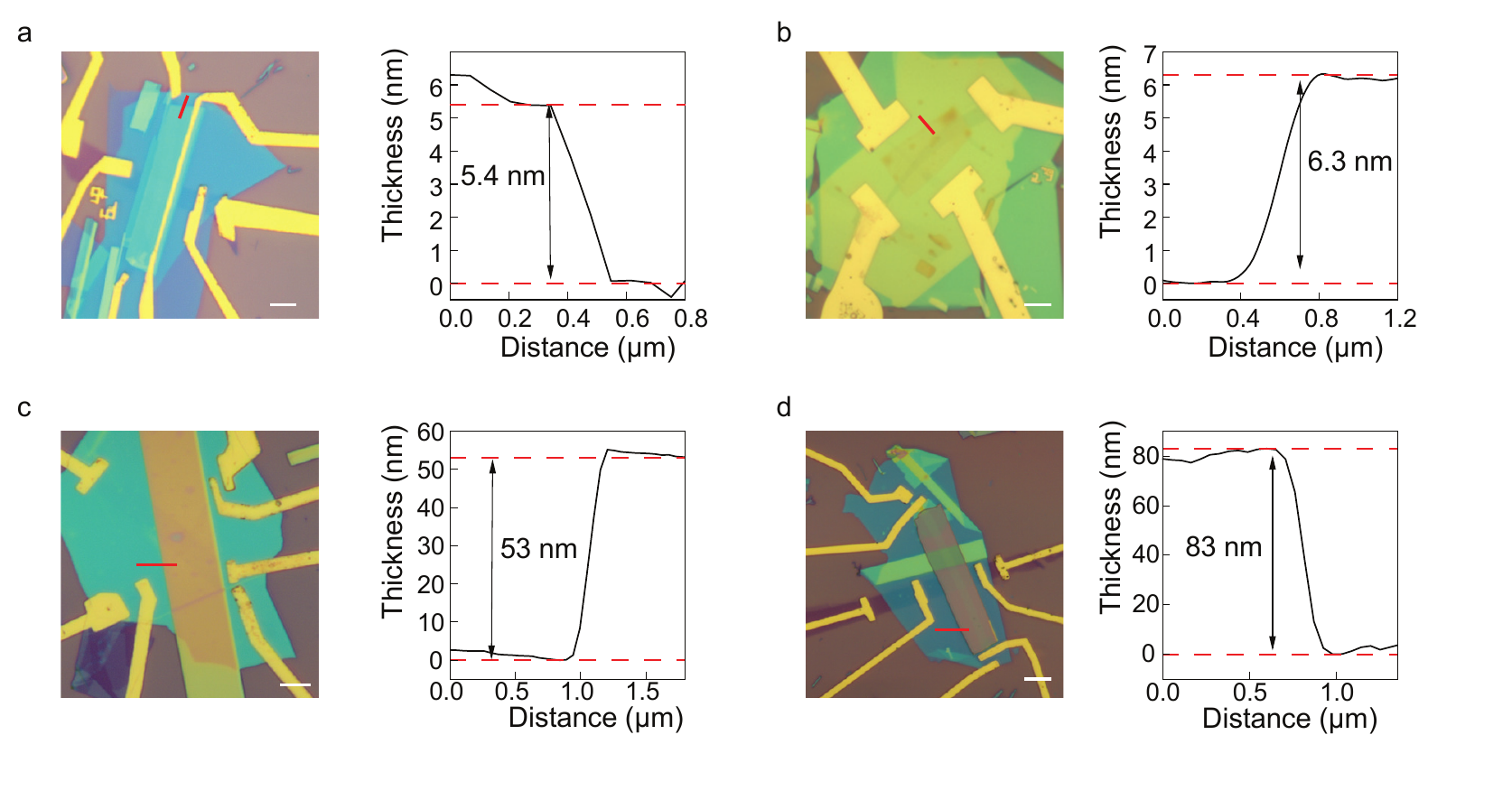}
  \caption{ Optical microscope images (left panels) and height profiles extracted from atomic force microscopy measurements (right panels; each profile is measured  along the red dashed line traces shown in the left panel) for devices A1/A2 (a), B (b), C (c), and D (d). In all microscope images the scale bar is 5 $\mu$m.} 
  \label{sifig1}
\end{figure*}
The vertical junction devices investigated in this work were fabricated using 5.4 nm (A1 and A2), 6.3 nm (B), 53 nm (C), and 83 nm (D) thick CrSBr crystals (see Fig \ref{sifig1} for atomic force microscope profiles), exfoliated from bulk crystals grown in-house by means of chemical vapor transport (see Ref.\cite{wu_quasi-1d_2022} for details on the crystal growth and characterization) . Although CrSBr does not exhibit signs of degradation when exposed to ambient conditions for short periods of time, the devices were assembled in a nitrogen gas-filled glove box with sub-ppm oxygen and water concentration, using now-established pick-up and transfer techniques (the substrates are made of silicon covered with 285 nm SiO$_2$). The device structure consists of exfoliated multilayer graphene top and bottom contacts placed over a bottom h-BN layer (in device B we also added a top h-BN layer to encapsulate the exfoliated CrSBr crystal completely, and no significant differences in the transport properties were observed as compared to the non-encapsulated devices). The Pt/Au leads connected to the multilayer graphene contacts were fabricated via standard e-beam lithography and evaporation followed by lift-off (in device B the connection to the multilayer graphene electrode was made via Cr/Au edge contacts using reactive ion etching with a CF$_4$/O$_2$ mixture prior to metal deposition).\\ 

Finally, junction A1 and junction A2 (see Fig.~\ref{sifig1} for optical images of the devices) are realized on a same CrSBr exfoliated crystals but with a different contact configuration. Specifically, the bottom graphene contact is shared by the two junctions but the top contact is different in the two cases: it consists of graphene for junction A1 and of a Pt film directly evaporated on top of CrSBr for junction A2.  Such a configuration allows us to compare two devices realized on a same CrSBr crystal with different contacts, which is useful to confirm that the contact resistance is not determining the behavior observed experimentally. Indeed, as discussed in section S4, the contact material has no influence on the linear regime of the current-voltage ($I$-$V$) characteristics of the devices.

\subsection{Electronic transport measurements}
Transport measurements were performed using a Teslatron Cryogen-free II cryostat (Oxford Instruments) equipped with a variable temperature insert, allowing us to stabilize the temperature at any value between  room temperature and 1.7~K range. A 12~T superconducting magnet was used to apply magnetic field. Electrical measurements were performed in a 4-terminal voltage bias configuration, using a home-made voltage source and low-noise current/voltage amplifiers connected to digital multimeters (Keysight 34401A).

\section{APPENDIX B: I-V characteristics of C\MakeLowercase{r}SB\MakeLowercase{r} multilayers}

\begin{figure}[t!]
  \includegraphics[width=1\linewidth]{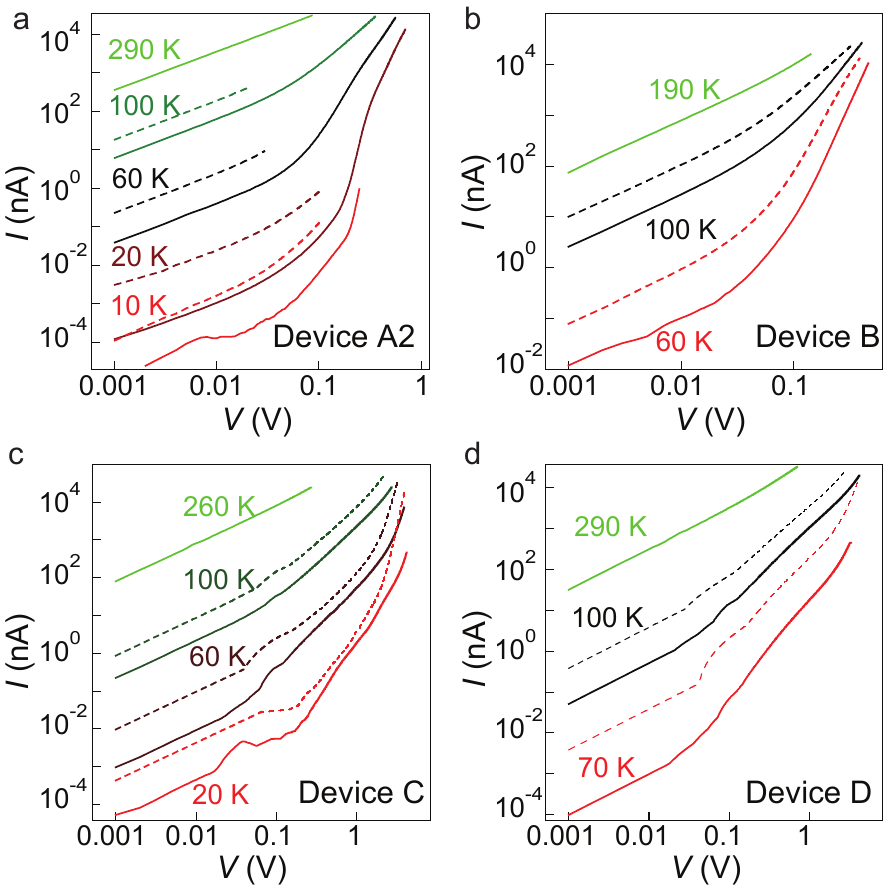}
  \caption{ $I$-$V$ curves of device A2 (a), B (b), C (c), and D (d) plotted in double-logarithmic scale, measured at different temperatures (as indicated next to each curve) and magnetic fields (the solid and dashed lines correspond to data measured at $\mu_0H = 0~T$ and $\mu_0H = 2~T$). } 
  \label{sifig2}
\end{figure}

\begin{figure}[t!]
  \includegraphics[width=1\linewidth]{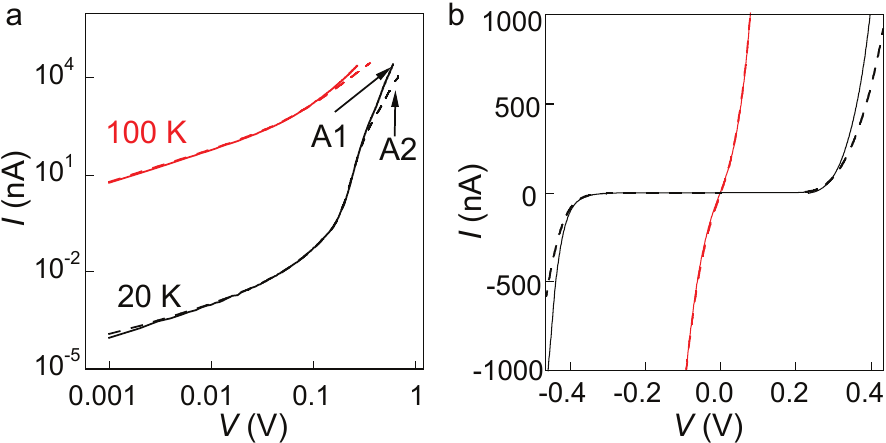}
  \caption{(a) Double logarithmic and (b) linear scale $I$-$V$ characteristic of graphene/CrSBr/graphene (A1) and graphite/CrSBr/Pt (A2) devices realized on the same CrSBr multilayer (the solid and dashed line represent data from device A1 and A2, respectively). Positive bias corresponds to injecting electrons from the top electrode (multilayer graphene for device A1 and Pt for device A2) and negative bias to injecting electrons from the common bottom multilayer graphene electrode. Data taken at zero applied magnetic field.} 
  \label{sifig3}
\end{figure}

As discussed in the main text, the investigation of thermally activated transport and variable range hopping (VRH) relies on the observation of the linear transport regime in the current-voltage ($I$-$V$) characteristics of our devices. In the main text we mostly show data for device A1. Here we show that the linear transport regime can be detected in all the investigated devices, as illustrated in Fig. \ref{sifig2}, which presents the $I$-$V$ characteristics of devices A2, B, C and D in a double-logarithmic scale.  Except for device D, whose resistance in the linear regime was too large to be measured below 40~K, and device B, which broke at 60~K, the linear regime was investigated down to 10 K, at both 0 T and 2 T (sufficient to bring CrSBr in the spin-flip phase, in which  all the spins are aligned parallel to the applied field). In summary, thermally activated transport was measured on all five devices realized and the variable range hopping regime was studied in three out of five devices. 

As mentioned above, we have also compared the $I$-$V$ characteristics of devices A1 and A2  realized on a same CrSBr crystal with different contacts, (see Section S2 for details), to ensure that the any effect of the contact resistance on our measurements is negligible (i.e., our experiments really probe the  electronic properties of CrSBr and not of the interface with the metal conacts). As can be seen in Fig \ref{sifig3} the $I$-$V$ curves measured in the two junctions perfectly fall  on top of each other in the linear regime, and start to deviate slightly only at larger applied bias, when the devices are biased outside the  the linear regime. This observation, together with the reproducibility of all measured quantities (resistivity, activation energy, $T_0$) in devices whose thickness ranges from 5 nm to nearly 100 nm fully confirm that our measurements of transport in the linear regime do probe the properties of CrSBr and are not affected by the contact resistance.

\section*{APPENDIX c: Analysis of activation energy ($E_a$) and characteristic temperature ($T_0$) for different devices}

As shown in the main text, the analysis of the resistance extracted from the linear regime revealed the presence of thermally activated behavior from 200 to 60 K, followed by VRH from 40 to 10 K, with the exact temperature range depending slightly on the investigated device. In the main text we show data from device A1. For completeness, here we show the data measured on all our devices, which we used to extract the average activation energy ($E_a$) and characteristics temperature ($T_0$) reported in the main text and summarized in Table \ref{tab1}.\\

\begin{figure}[htbp]
  \includegraphics[width=1\linewidth]{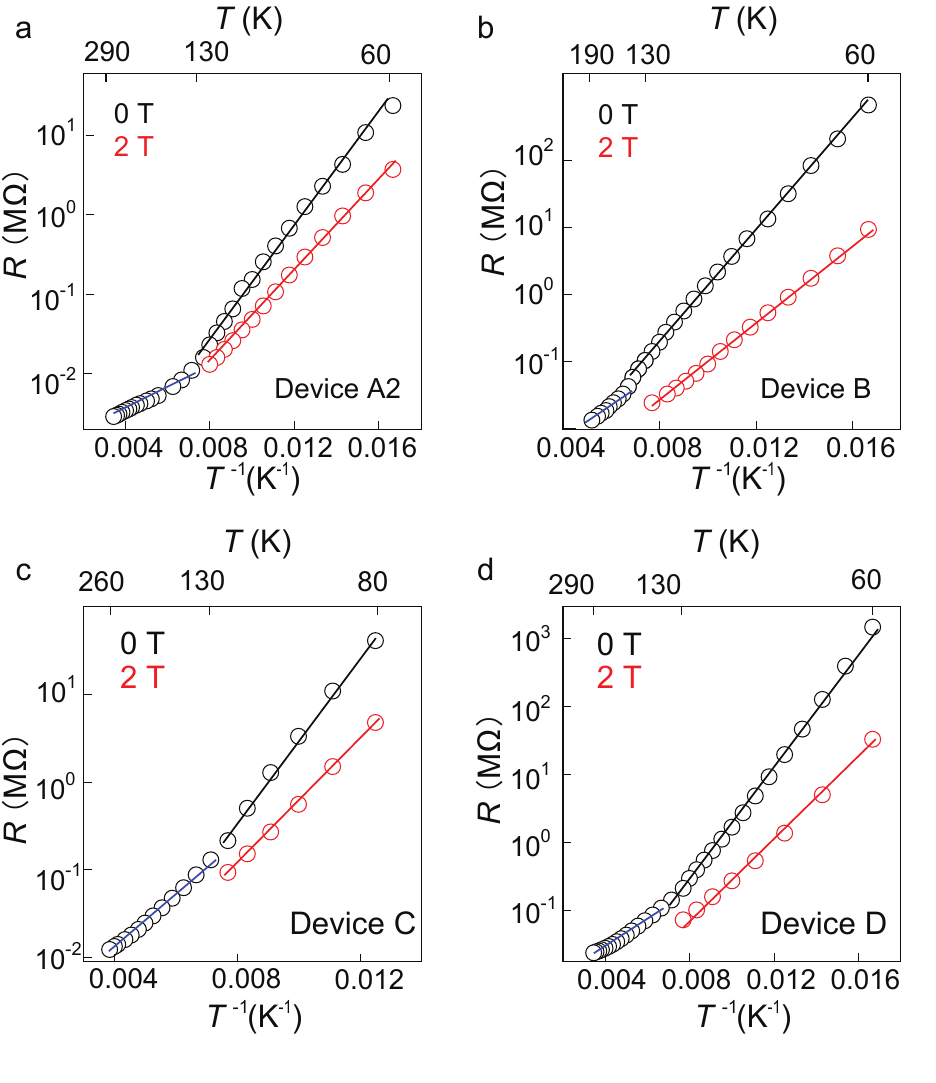}
  \caption{Arrhenius plot of resistance as a function of inverse temperature for devices A2 (a), B(b), C(c), D(d). The black, red, and blue lines represent linear fits to the data in the antiferromagnetic, spin-flip, and paramagnetic states respectively.} 
  \label{sifig4}
\end{figure}

\begin{figure}[t!]
  \includegraphics[width=1\linewidth]{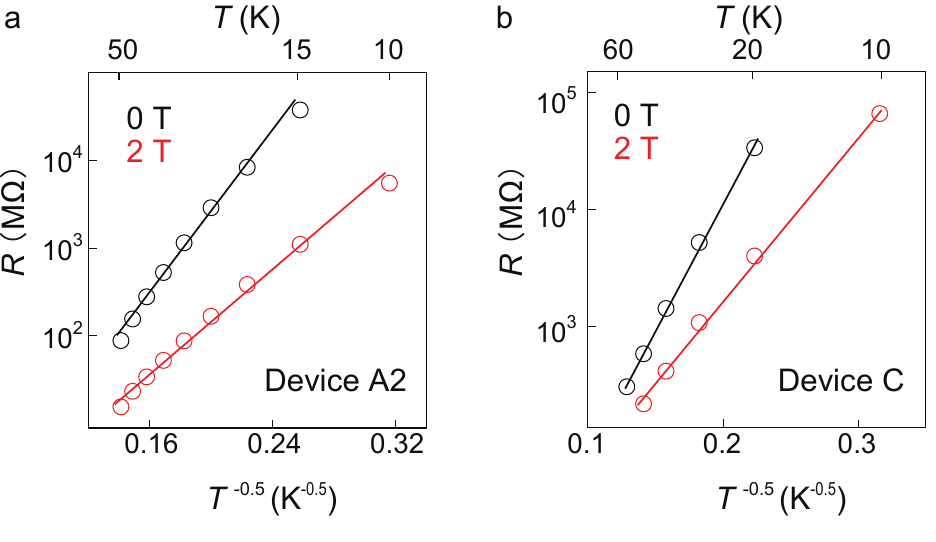}
  \caption{(a), (b) Semi-logarithmic plot of the resistance versus ${T}^{-0.5}$ for devices A2 and C. The black and red line represents linear fits to the data with applied magnetic field  $\mu_0H =$ 0 and 2~T.} 
  \label{sifig5}
\end{figure}

Fig. \ref{sifig4} shows the Arrhenius plot obtained for devices A2, B, C and D in the paramagnetic (~290 K to 140 K; 0~T), antiferromagnetic (~130 K to 60 K; 0~T) and spin-flip (~130 K to 60 K; 2~T) phase. Fig. \ref{sifig5} shows a plot of $\ln(R)$ and $1/T^{1/2}$ (corresponding to the Efros-Skhlovskii VRH regime) between 50 and 10 K in the antiferromagnetic state (0~T) and spin-flip phase (2~T) for devices A2 and C. The values of $E_a$ extracted from linear fits of $ln(R)$ vs $1/T$ as well as the values of $T_0$ extracted from linear fits of $ln(R)$ vs $1/T^{1/2}$ in the different magnetic states are summarized in Table \ref{tab1}.

 \section{APPENDIX D: Magnetotransport in the linear transport regime}
\begin{figure}[htbp]
   \includegraphics[width=1\linewidth]{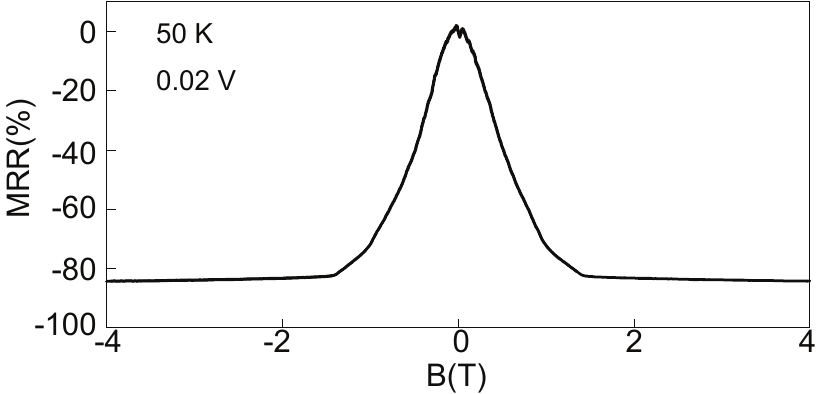}
   \caption{Magnetoresistance $(R(B)-R(B=0))/R(B=0)$ measured on device A2 at $T=50$ K with an applied bias of $V=0.02$ V (corresponding to having the device biased in the linear part of the $I-V$ curve). The  magnetic field is parallel to the c-axis. .} 
  \label{sifig6}
\end{figure}

 In the main text, we show the analysis of temperature-dependent resistance both at  $\mu_0H = 0~T$ (i.e., in the antiferromagnetic state of CrSBr) and at  $\mu_0H = 2~T$ (i.e., in the spin-flip state, with all spin pointing in the same direction). For completeness, here we show  magnetoresistance data measured at  50 K in  linear part of the $I$-$V$ curve, upon varying the applied perpendicular magnetic field from -4 T to 4 T. As shown in Fig \ref{sifig6}, a continuously varying magnetoresistance $(R(B)-R(B=0))/R(B=0)$ is observed, whose magnitude correspond approximately to a factor of 2 change in resistance. The magnetoresistance is clearly seen to saturate when the applied magnetic field exceed the spin-flip field (close to 1.5 T). Finding that the resistance decreases upon applying magnetic field to align the magnetization in all of layers is a typical for vertical transport through A-type antiferromagnetic semiconductors, irrespective of the transport regime.

\FloatBarrier

\begin{thebibliography}{38}%
\makeatletter
\providecommand \@ifxundefined [1]{%
 \@ifx{#1\undefined}
}%
\providecommand \@ifnum [1]{%
 \ifnum #1\expandafter \@firstoftwo
 \else \expandafter \@secondoftwo
 \fi
}%
\providecommand \@ifx [1]{%
 \ifx #1\expandafter \@firstoftwo
 \else \expandafter \@secondoftwo
 \fi
}%
\providecommand \natexlab [1]{#1}%
\providecommand \enquote  [1]{``#1''}%
\providecommand \bibnamefont  [1]{#1}%
\providecommand \bibfnamefont [1]{#1}%
\providecommand \citenamefont [1]{#1}%
\providecommand \href@noop [0]{\@secondoftwo}%
\providecommand \href [0]{\begingroup \@sanitize@url \@href}%
\providecommand \@href[1]{\@@startlink{#1}\@@href}%
\providecommand \@@href[1]{\endgroup#1\@@endlink}%
\providecommand \@sanitize@url [0]{\catcode `\\12\catcode `\$12\catcode `\&12\catcode `\#12\catcode `\^12\catcode `\_12\catcode `\%12\relax}%
\providecommand \@@startlink[1]{}%
\providecommand \@@endlink[0]{}%
\providecommand \url  [0]{\begingroup\@sanitize@url \@url }%
\providecommand \@url [1]{\endgroup\@href {#1}{\urlprefix }}%
\providecommand \urlprefix  [0]{URL }%
\providecommand \Eprint [0]{\href }%
\providecommand \doibase [0]{https://doi.org/}%
\providecommand \selectlanguage [0]{\@gobble}%
\providecommand \bibinfo  [0]{\@secondoftwo}%
\providecommand \bibfield  [0]{\@secondoftwo}%
\providecommand \translation [1]{[#1]}%
\providecommand \BibitemOpen [0]{}%
\providecommand \bibitemStop [0]{}%
\providecommand \bibitemNoStop [0]{.\EOS\space}%
\providecommand \EOS [0]{\spacefactor3000\relax}%
\providecommand \BibitemShut  [1]{\csname bibitem#1\endcsname}%
\let\auto@bib@innerbib\@empty
\bibitem [{\citenamefont {Burch}\ \emph {et~al.}(2018)\citenamefont {Burch}, \citenamefont {Mandrus},\ and\ \citenamefont {Park}}]{burch_magnetism_2018}%
  \BibitemOpen
  \bibfield  {author} {\bibinfo {author} {\bibfnamefont {K.~S.}\ \bibnamefont {Burch}}, \bibinfo {author} {\bibfnamefont {D.}~\bibnamefont {Mandrus}},\ and\ \bibinfo {author} {\bibfnamefont {J.-G.}\ \bibnamefont {Park}},\ }\bibfield  {title} {\bibinfo {title} {Magnetism in two-dimensional van der {Waals} materials},\ }\href {https://doi.org/10.1038/s41586-018-0631-z} {\bibfield  {journal} {\bibinfo  {journal} {Nature}\ }\textbf {\bibinfo {volume} {563}},\ \bibinfo {pages} {47} (\bibinfo {year} {2018})}\BibitemShut {NoStop}%
\bibitem [{\citenamefont {Gibertini}\ \emph {et~al.}(2019)\citenamefont {Gibertini}, \citenamefont {Koperski}, \citenamefont {Morpurgo},\ and\ \citenamefont {Novoselov}}]{gibertini_magnetic_2019}%
  \BibitemOpen
  \bibfield  {author} {\bibinfo {author} {\bibfnamefont {M.}~\bibnamefont {Gibertini}}, \bibinfo {author} {\bibfnamefont {M.}~\bibnamefont {Koperski}}, \bibinfo {author} {\bibfnamefont {A.~F.}\ \bibnamefont {Morpurgo}},\ and\ \bibinfo {author} {\bibfnamefont {K.~S.}\ \bibnamefont {Novoselov}},\ }\bibfield  {title} {\bibinfo {title} {Magnetic {2D} materials and heterostructures},\ }\href {https://doi.org/10.1038/s41565-019-0438-6} {\bibfield  {journal} {\bibinfo  {journal} {Nature Nanotechnology}\ }\textbf {\bibinfo {volume} {14}},\ \bibinfo {pages} {408} (\bibinfo {year} {2019})}\BibitemShut {NoStop}%
\bibitem [{\citenamefont {Mak}\ \emph {et~al.}(2019)\citenamefont {Mak}, \citenamefont {Shan},\ and\ \citenamefont {Ralph}}]{mak_probing_2019}%
  \BibitemOpen
  \bibfield  {author} {\bibinfo {author} {\bibfnamefont {K.~F.}\ \bibnamefont {Mak}}, \bibinfo {author} {\bibfnamefont {J.}~\bibnamefont {Shan}},\ and\ \bibinfo {author} {\bibfnamefont {D.~C.}\ \bibnamefont {Ralph}},\ }\bibfield  {title} {\bibinfo {title} {Probing and controlling magnetic states in {2D} layered magnetic materials},\ }\href {https://doi.org/10.1038/s42254-019-0110-y} {\bibfield  {journal} {\bibinfo  {journal} {Nature Reviews Physics}\ }\textbf {\bibinfo {volume} {1}},\ \bibinfo {pages} {646} (\bibinfo {year} {2019})}\BibitemShut {NoStop}%
\bibitem [{\citenamefont {Wang}\ \emph {et~al.}(2019)\citenamefont {Wang}, \citenamefont {Gibertini}, \citenamefont {Dumcenco}, \citenamefont {Taniguchi}, \citenamefont {Watanabe}, \citenamefont {Giannini},\ and\ \citenamefont {Morpurgo}}]{wang_determining_2019}%
  \BibitemOpen
  \bibfield  {author} {\bibinfo {author} {\bibfnamefont {Z.}~\bibnamefont {Wang}}, \bibinfo {author} {\bibfnamefont {M.}~\bibnamefont {Gibertini}}, \bibinfo {author} {\bibfnamefont {D.}~\bibnamefont {Dumcenco}}, \bibinfo {author} {\bibfnamefont {T.}~\bibnamefont {Taniguchi}}, \bibinfo {author} {\bibfnamefont {K.}~\bibnamefont {Watanabe}}, \bibinfo {author} {\bibfnamefont {E.}~\bibnamefont {Giannini}},\ and\ \bibinfo {author} {\bibfnamefont {A.~F.}\ \bibnamefont {Morpurgo}},\ }\bibfield  {title} {\bibinfo {title} {Determining the phase diagram of atomically thin layered antiferromagnet {CrCl3}},\ }\href {https://doi.org/10.1038/s41565-019-0565-0} {\bibfield  {journal} {\bibinfo  {journal} {Nature Nanotechnology}\ }\textbf {\bibinfo {volume} {14}},\ \bibinfo {pages} {1116} (\bibinfo {year} {2019})}\BibitemShut {NoStop}%
\bibitem [{\citenamefont {Long}\ \emph {et~al.}(2020)\citenamefont {Long}, \citenamefont {Henck}, \citenamefont {Gibertini}, \citenamefont {Dumcenco}, \citenamefont {Wang}, \citenamefont {Taniguchi}, \citenamefont {Watanabe}, \citenamefont {Giannini},\ and\ \citenamefont {Morpurgo}}]{long_persistence_2020}%
  \BibitemOpen
  \bibfield  {author} {\bibinfo {author} {\bibfnamefont {G.}~\bibnamefont {Long}}, \bibinfo {author} {\bibfnamefont {H.}~\bibnamefont {Henck}}, \bibinfo {author} {\bibfnamefont {M.}~\bibnamefont {Gibertini}}, \bibinfo {author} {\bibfnamefont {D.}~\bibnamefont {Dumcenco}}, \bibinfo {author} {\bibfnamefont {Z.}~\bibnamefont {Wang}}, \bibinfo {author} {\bibfnamefont {T.}~\bibnamefont {Taniguchi}}, \bibinfo {author} {\bibfnamefont {K.}~\bibnamefont {Watanabe}}, \bibinfo {author} {\bibfnamefont {E.}~\bibnamefont {Giannini}},\ and\ \bibinfo {author} {\bibfnamefont {A.~F.}\ \bibnamefont {Morpurgo}},\ }\bibfield  {title} {\bibinfo {title} {Persistence of {Magnetism} in {Atomically} {Thin} {MnPS} $_{\textrm{3}}$ {Crystals}},\ }\href {https://doi.org/10.1021/acs.nanolett.9b05165} {\bibfield  {journal} {\bibinfo  {journal} {Nano Letters}\ }\textbf {\bibinfo {volume} {20}},\ \bibinfo {pages} {2452} (\bibinfo {year} {2020})}\BibitemShut {NoStop}%
\bibitem [{\citenamefont {Gong}\ \emph {et~al.}(2017)\citenamefont {Gong}, \citenamefont {Li}, \citenamefont {Li}, \citenamefont {Ji}, \citenamefont {Stern}, \citenamefont {Xia}, \citenamefont {Cao}, \citenamefont {Bao}, \citenamefont {Wang}, \citenamefont {Wang}, \citenamefont {Qiu}, \citenamefont {Cava}, \citenamefont {Louie}, \citenamefont {Xia},\ and\ \citenamefont {Zhang}}]{gong_discovery_2017}%
  \BibitemOpen
  \bibfield  {author} {\bibinfo {author} {\bibfnamefont {C.}~\bibnamefont {Gong}}, \bibinfo {author} {\bibfnamefont {L.}~\bibnamefont {Li}}, \bibinfo {author} {\bibfnamefont {Z.}~\bibnamefont {Li}}, \bibinfo {author} {\bibfnamefont {H.}~\bibnamefont {Ji}}, \bibinfo {author} {\bibfnamefont {A.}~\bibnamefont {Stern}}, \bibinfo {author} {\bibfnamefont {Y.}~\bibnamefont {Xia}}, \bibinfo {author} {\bibfnamefont {T.}~\bibnamefont {Cao}}, \bibinfo {author} {\bibfnamefont {W.}~\bibnamefont {Bao}}, \bibinfo {author} {\bibfnamefont {C.}~\bibnamefont {Wang}}, \bibinfo {author} {\bibfnamefont {Y.}~\bibnamefont {Wang}}, \bibinfo {author} {\bibfnamefont {Z.~Q.}\ \bibnamefont {Qiu}}, \bibinfo {author} {\bibfnamefont {R.~J.}\ \bibnamefont {Cava}}, \bibinfo {author} {\bibfnamefont {S.~G.}\ \bibnamefont {Louie}}, \bibinfo {author} {\bibfnamefont {J.}~\bibnamefont {Xia}},\ and\ \bibinfo {author} {\bibfnamefont {X.}~\bibnamefont {Zhang}},\ }\bibfield  {title} {\bibinfo {title} {Discovery of intrinsic ferromagnetism in
  two-dimensional van der {Waals} crystals},\ }\href {https://doi.org/10.1038/nature22060} {\bibfield  {journal} {\bibinfo  {journal} {Nature}\ }\textbf {\bibinfo {volume} {546}},\ \bibinfo {pages} {265} (\bibinfo {year} {2017})}\BibitemShut {NoStop}%
\bibitem [{\citenamefont {Huang}\ \emph {et~al.}(2017)\citenamefont {Huang}, \citenamefont {Clark}, \citenamefont {Navarro-Moratalla}, \citenamefont {Klein}, \citenamefont {Cheng}, \citenamefont {Seyler}, \citenamefont {Zhong}, \citenamefont {Schmidgall}, \citenamefont {McGuire}, \citenamefont {Cobden}, \citenamefont {Yao}, \citenamefont {Xiao}, \citenamefont {Jarillo-Herrero},\ and\ \citenamefont {Xu}}]{huang_layer-dependent_2017}%
  \BibitemOpen
  \bibfield  {author} {\bibinfo {author} {\bibfnamefont {B.}~\bibnamefont {Huang}}, \bibinfo {author} {\bibfnamefont {G.}~\bibnamefont {Clark}}, \bibinfo {author} {\bibfnamefont {E.}~\bibnamefont {Navarro-Moratalla}}, \bibinfo {author} {\bibfnamefont {D.~R.}\ \bibnamefont {Klein}}, \bibinfo {author} {\bibfnamefont {R.}~\bibnamefont {Cheng}}, \bibinfo {author} {\bibfnamefont {K.~L.}\ \bibnamefont {Seyler}}, \bibinfo {author} {\bibfnamefont {D.}~\bibnamefont {Zhong}}, \bibinfo {author} {\bibfnamefont {E.}~\bibnamefont {Schmidgall}}, \bibinfo {author} {\bibfnamefont {M.~A.}\ \bibnamefont {McGuire}}, \bibinfo {author} {\bibfnamefont {D.~H.}\ \bibnamefont {Cobden}}, \bibinfo {author} {\bibfnamefont {W.}~\bibnamefont {Yao}}, \bibinfo {author} {\bibfnamefont {D.}~\bibnamefont {Xiao}}, \bibinfo {author} {\bibfnamefont {P.}~\bibnamefont {Jarillo-Herrero}},\ and\ \bibinfo {author} {\bibfnamefont {X.}~\bibnamefont {Xu}},\ }\bibfield  {title} {\bibinfo {title} {Layer-dependent ferromagnetism in a van der {Waals} crystal
  down to the monolayer limit},\ }\href {https://doi.org/10.1038/nature22391} {\bibfield  {journal} {\bibinfo  {journal} {Nature}\ }\textbf {\bibinfo {volume} {546}},\ \bibinfo {pages} {270} (\bibinfo {year} {2017})}\BibitemShut {NoStop}%
\bibitem [{\citenamefont {Klein}\ \emph {et~al.}(2018)\citenamefont {Klein}, \citenamefont {MacNeill}, \citenamefont {Lado}, \citenamefont {Soriano}, \citenamefont {Navarro-Moratalla}, \citenamefont {Watanabe}, \citenamefont {Taniguchi}, \citenamefont {Manni}, \citenamefont {Canfield}, \citenamefont {Fernández-Rossier},\ and\ \citenamefont {Jarillo-Herrero}}]{klein_probing_2018}%
  \BibitemOpen
  \bibfield  {author} {\bibinfo {author} {\bibfnamefont {D.~R.}\ \bibnamefont {Klein}}, \bibinfo {author} {\bibfnamefont {D.}~\bibnamefont {MacNeill}}, \bibinfo {author} {\bibfnamefont {J.~L.}\ \bibnamefont {Lado}}, \bibinfo {author} {\bibfnamefont {D.}~\bibnamefont {Soriano}}, \bibinfo {author} {\bibfnamefont {E.}~\bibnamefont {Navarro-Moratalla}}, \bibinfo {author} {\bibfnamefont {K.}~\bibnamefont {Watanabe}}, \bibinfo {author} {\bibfnamefont {T.}~\bibnamefont {Taniguchi}}, \bibinfo {author} {\bibfnamefont {S.}~\bibnamefont {Manni}}, \bibinfo {author} {\bibfnamefont {P.}~\bibnamefont {Canfield}}, \bibinfo {author} {\bibfnamefont {J.}~\bibnamefont {Fernández-Rossier}},\ and\ \bibinfo {author} {\bibfnamefont {P.}~\bibnamefont {Jarillo-Herrero}},\ }\bibfield  {title} {\bibinfo {title} {Probing magnetism in {2D} van der {Waals} crystalline insulators via electron tunneling},\ }\href {https://doi.org/10.1126/science.aar3617} {\bibfield  {journal} {\bibinfo  {journal} {Science}\ }\textbf {\bibinfo {volume}
  {360}},\ \bibinfo {pages} {1218} (\bibinfo {year} {2018})}\BibitemShut {NoStop}%
\bibitem [{\citenamefont {Song}\ \emph {et~al.}(2018)\citenamefont {Song}, \citenamefont {Cai}, \citenamefont {Tu}, \citenamefont {Zhang}, \citenamefont {Huang}, \citenamefont {Wilson}, \citenamefont {Seyler}, \citenamefont {Zhu}, \citenamefont {Taniguchi}, \citenamefont {Watanabe}, \citenamefont {McGuire}, \citenamefont {Cobden}, \citenamefont {Xiao}, \citenamefont {Yao},\ and\ \citenamefont {Xu}}]{song_giant_2018}%
  \BibitemOpen
  \bibfield  {author} {\bibinfo {author} {\bibfnamefont {T.}~\bibnamefont {Song}}, \bibinfo {author} {\bibfnamefont {X.}~\bibnamefont {Cai}}, \bibinfo {author} {\bibfnamefont {M.~W.-Y.}\ \bibnamefont {Tu}}, \bibinfo {author} {\bibfnamefont {X.}~\bibnamefont {Zhang}}, \bibinfo {author} {\bibfnamefont {B.}~\bibnamefont {Huang}}, \bibinfo {author} {\bibfnamefont {N.~P.}\ \bibnamefont {Wilson}}, \bibinfo {author} {\bibfnamefont {K.~L.}\ \bibnamefont {Seyler}}, \bibinfo {author} {\bibfnamefont {L.}~\bibnamefont {Zhu}}, \bibinfo {author} {\bibfnamefont {T.}~\bibnamefont {Taniguchi}}, \bibinfo {author} {\bibfnamefont {K.}~\bibnamefont {Watanabe}}, \bibinfo {author} {\bibfnamefont {M.~A.}\ \bibnamefont {McGuire}}, \bibinfo {author} {\bibfnamefont {D.~H.}\ \bibnamefont {Cobden}}, \bibinfo {author} {\bibfnamefont {D.}~\bibnamefont {Xiao}}, \bibinfo {author} {\bibfnamefont {W.}~\bibnamefont {Yao}},\ and\ \bibinfo {author} {\bibfnamefont {X.}~\bibnamefont {Xu}},\ }\bibfield  {title} {\bibinfo {title} {Giant tunneling
  magnetoresistance in spin-filter van der {Waals} heterostructures},\ }\href {https://doi.org/10.1126/science.aar4851} {\bibfield  {journal} {\bibinfo  {journal} {Science}\ }\textbf {\bibinfo {volume} {360}},\ \bibinfo {pages} {1214} (\bibinfo {year} {2018})}\BibitemShut {NoStop}%
\bibitem [{\citenamefont {Kim}\ \emph {et~al.}(2018)\citenamefont {Kim}, \citenamefont {Yang}, \citenamefont {Patel}, \citenamefont {Sfigakis}, \citenamefont {Li}, \citenamefont {Tian}, \citenamefont {Lei},\ and\ \citenamefont {Tsen}}]{kim_one_2018}%
  \BibitemOpen
  \bibfield  {author} {\bibinfo {author} {\bibfnamefont {H.~H.}\ \bibnamefont {Kim}}, \bibinfo {author} {\bibfnamefont {B.}~\bibnamefont {Yang}}, \bibinfo {author} {\bibfnamefont {T.}~\bibnamefont {Patel}}, \bibinfo {author} {\bibfnamefont {F.}~\bibnamefont {Sfigakis}}, \bibinfo {author} {\bibfnamefont {C.}~\bibnamefont {Li}}, \bibinfo {author} {\bibfnamefont {S.}~\bibnamefont {Tian}}, \bibinfo {author} {\bibfnamefont {H.}~\bibnamefont {Lei}},\ and\ \bibinfo {author} {\bibfnamefont {A.~W.}\ \bibnamefont {Tsen}},\ }\bibfield  {title} {\bibinfo {title} {One {Million} {Percent} {Tunnel} {Magnetoresistance} in a {Magnetic} van der {Waals} {Heterostructure}},\ }\href {https://doi.org/10.1021/acs.nanolett.8b01552} {\bibfield  {journal} {\bibinfo  {journal} {Nano Letters}\ }\textbf {\bibinfo {volume} {18}},\ \bibinfo {pages} {4885} (\bibinfo {year} {2018})}\BibitemShut {NoStop}%
\bibitem [{\citenamefont {Wang}\ \emph {et~al.}(2018)\citenamefont {Wang}, \citenamefont {Zhang}, \citenamefont {Ding}, \citenamefont {Dong}, \citenamefont {Li}, \citenamefont {Chen}, \citenamefont {Li}, \citenamefont {Huang}, \citenamefont {Wang}, \citenamefont {Zhao}, \citenamefont {Li}, \citenamefont {Li}, \citenamefont {Jia}, \citenamefont {Sun}, \citenamefont {Guo}, \citenamefont {Ye}, \citenamefont {Sun}, \citenamefont {Chen}, \citenamefont {Yang}, \citenamefont {Zhang}, \citenamefont {Ono}, \citenamefont {Han},\ and\ \citenamefont {Zhang}}]{wang_electric-field_2018}%
  \BibitemOpen
  \bibfield  {author} {\bibinfo {author} {\bibfnamefont {Z.}~\bibnamefont {Wang}}, \bibinfo {author} {\bibfnamefont {T.}~\bibnamefont {Zhang}}, \bibinfo {author} {\bibfnamefont {M.}~\bibnamefont {Ding}}, \bibinfo {author} {\bibfnamefont {B.}~\bibnamefont {Dong}}, \bibinfo {author} {\bibfnamefont {Y.}~\bibnamefont {Li}}, \bibinfo {author} {\bibfnamefont {M.}~\bibnamefont {Chen}}, \bibinfo {author} {\bibfnamefont {X.}~\bibnamefont {Li}}, \bibinfo {author} {\bibfnamefont {J.}~\bibnamefont {Huang}}, \bibinfo {author} {\bibfnamefont {H.}~\bibnamefont {Wang}}, \bibinfo {author} {\bibfnamefont {X.}~\bibnamefont {Zhao}}, \bibinfo {author} {\bibfnamefont {Y.}~\bibnamefont {Li}}, \bibinfo {author} {\bibfnamefont {D.}~\bibnamefont {Li}}, \bibinfo {author} {\bibfnamefont {C.}~\bibnamefont {Jia}}, \bibinfo {author} {\bibfnamefont {L.}~\bibnamefont {Sun}}, \bibinfo {author} {\bibfnamefont {H.}~\bibnamefont {Guo}}, \bibinfo {author} {\bibfnamefont {Y.}~\bibnamefont {Ye}}, \bibinfo {author} {\bibfnamefont {D.}~\bibnamefont
  {Sun}}, \bibinfo {author} {\bibfnamefont {Y.}~\bibnamefont {Chen}}, \bibinfo {author} {\bibfnamefont {T.}~\bibnamefont {Yang}}, \bibinfo {author} {\bibfnamefont {J.}~\bibnamefont {Zhang}}, \bibinfo {author} {\bibfnamefont {S.}~\bibnamefont {Ono}}, \bibinfo {author} {\bibfnamefont {Z.}~\bibnamefont {Han}},\ and\ \bibinfo {author} {\bibfnamefont {Z.}~\bibnamefont {Zhang}},\ }\bibfield  {title} {\bibinfo {title} {Electric-field control of magnetism in a few-layered van der {Waals} ferromagnetic semiconductor},\ }\href {https://doi.org/10.1038/s41565-018-0186-z} {\bibfield  {journal} {\bibinfo  {journal} {Nature Nanotechnology}\ }\textbf {\bibinfo {volume} {13}},\ \bibinfo {pages} {554} (\bibinfo {year} {2018})}\BibitemShut {NoStop}%
\bibitem [{\citenamefont {Kim}\ \emph {et~al.}(2019)\citenamefont {Kim}, \citenamefont {Yang}, \citenamefont {Tian}, \citenamefont {Li}, \citenamefont {Miao}, \citenamefont {Lei},\ and\ \citenamefont {Tsen}}]{kim_tailored_2019}%
  \BibitemOpen
  \bibfield  {author} {\bibinfo {author} {\bibfnamefont {H.~H.}\ \bibnamefont {Kim}}, \bibinfo {author} {\bibfnamefont {B.}~\bibnamefont {Yang}}, \bibinfo {author} {\bibfnamefont {S.}~\bibnamefont {Tian}}, \bibinfo {author} {\bibfnamefont {C.}~\bibnamefont {Li}}, \bibinfo {author} {\bibfnamefont {G.-X.}\ \bibnamefont {Miao}}, \bibinfo {author} {\bibfnamefont {H.}~\bibnamefont {Lei}},\ and\ \bibinfo {author} {\bibfnamefont {A.~W.}\ \bibnamefont {Tsen}},\ }\bibfield  {title} {\bibinfo {title} {Tailored {Tunnel} {Magnetoresistance} {Response} in {Three} {Ultrathin} {Chromium} {Trihalides}},\ }\href {https://doi.org/10.1021/acs.nanolett.9b02357} {\bibfield  {journal} {\bibinfo  {journal} {Nano Letters}\ }\textbf {\bibinfo {volume} {19}},\ \bibinfo {pages} {5739} (\bibinfo {year} {2019})}\BibitemShut {NoStop}%
\bibitem [{\citenamefont {Song}\ \emph {et~al.}(2019)\citenamefont {Song}, \citenamefont {Fei}, \citenamefont {Yankowitz}, \citenamefont {Lin}, \citenamefont {Jiang}, \citenamefont {Hwangbo}, \citenamefont {Zhang}, \citenamefont {Sun}, \citenamefont {Taniguchi}, \citenamefont {Watanabe}, \citenamefont {McGuire}, \citenamefont {Graf}, \citenamefont {Cao}, \citenamefont {Chu}, \citenamefont {Cobden}, \citenamefont {Dean}, \citenamefont {Xiao},\ and\ \citenamefont {Xu}}]{song_switching_2019}%
  \BibitemOpen
  \bibfield  {author} {\bibinfo {author} {\bibfnamefont {T.}~\bibnamefont {Song}}, \bibinfo {author} {\bibfnamefont {Z.}~\bibnamefont {Fei}}, \bibinfo {author} {\bibfnamefont {M.}~\bibnamefont {Yankowitz}}, \bibinfo {author} {\bibfnamefont {Z.}~\bibnamefont {Lin}}, \bibinfo {author} {\bibfnamefont {Q.}~\bibnamefont {Jiang}}, \bibinfo {author} {\bibfnamefont {K.}~\bibnamefont {Hwangbo}}, \bibinfo {author} {\bibfnamefont {Q.}~\bibnamefont {Zhang}}, \bibinfo {author} {\bibfnamefont {B.}~\bibnamefont {Sun}}, \bibinfo {author} {\bibfnamefont {T.}~\bibnamefont {Taniguchi}}, \bibinfo {author} {\bibfnamefont {K.}~\bibnamefont {Watanabe}}, \bibinfo {author} {\bibfnamefont {M.~A.}\ \bibnamefont {McGuire}}, \bibinfo {author} {\bibfnamefont {D.}~\bibnamefont {Graf}}, \bibinfo {author} {\bibfnamefont {T.}~\bibnamefont {Cao}}, \bibinfo {author} {\bibfnamefont {J.-H.}\ \bibnamefont {Chu}}, \bibinfo {author} {\bibfnamefont {D.~H.}\ \bibnamefont {Cobden}}, \bibinfo {author} {\bibfnamefont {C.~R.}\ \bibnamefont {Dean}},
  \bibinfo {author} {\bibfnamefont {D.}~\bibnamefont {Xiao}},\ and\ \bibinfo {author} {\bibfnamefont {X.}~\bibnamefont {Xu}},\ }\bibfield  {title} {\bibinfo {title} {Switching {2D} magnetic states via pressure tuning of layer stacking},\ }\href {https://doi.org/10.1038/s41563-019-0505-2} {\bibfield  {journal} {\bibinfo  {journal} {Nature Materials}\ }\textbf {\bibinfo {volume} {18}},\ \bibinfo {pages} {1298} (\bibinfo {year} {2019})}\BibitemShut {NoStop}%
\bibitem [{\citenamefont {Wang}\ \emph {et~al.}(2021)\citenamefont {Wang}, \citenamefont {Gutiérrez-Lezama}, \citenamefont {Dumcenco}, \citenamefont {Ubrig}, \citenamefont {Taniguchi}, \citenamefont {Watanabe}, \citenamefont {Giannini}, \citenamefont {Gibertini},\ and\ \citenamefont {Morpurgo}}]{wang_magnetization_2021}%
  \BibitemOpen
  \bibfield  {author} {\bibinfo {author} {\bibfnamefont {Z.}~\bibnamefont {Wang}}, \bibinfo {author} {\bibfnamefont {I.}~\bibnamefont {Gutiérrez-Lezama}}, \bibinfo {author} {\bibfnamefont {D.}~\bibnamefont {Dumcenco}}, \bibinfo {author} {\bibfnamefont {N.}~\bibnamefont {Ubrig}}, \bibinfo {author} {\bibfnamefont {T.}~\bibnamefont {Taniguchi}}, \bibinfo {author} {\bibfnamefont {K.}~\bibnamefont {Watanabe}}, \bibinfo {author} {\bibfnamefont {E.}~\bibnamefont {Giannini}}, \bibinfo {author} {\bibfnamefont {M.}~\bibnamefont {Gibertini}},\ and\ \bibinfo {author} {\bibfnamefont {A.~F.}\ \bibnamefont {Morpurgo}},\ }\bibfield  {title} {\bibinfo {title} {Magnetization dependent tunneling conductance of ferromagnetic barriers},\ }\href {https://doi.org/10.1038/s41467-021-26973-7} {\bibfield  {journal} {\bibinfo  {journal} {Nature Communications}\ }\textbf {\bibinfo {volume} {12}},\ \bibinfo {pages} {6659} (\bibinfo {year} {2021})}\BibitemShut {NoStop}%
\bibitem [{\citenamefont {Soler-Delgado}\ \emph {et~al.}(2022)\citenamefont {Soler-Delgado}, \citenamefont {Yao}, \citenamefont {Dumcenco}, \citenamefont {Giannini}, \citenamefont {Li}, \citenamefont {Occhialini}, \citenamefont {Comin}, \citenamefont {Ubrig},\ and\ \citenamefont {Morpurgo}}]{soler-delgado_probing_2022}%
  \BibitemOpen
  \bibfield  {author} {\bibinfo {author} {\bibfnamefont {D.}~\bibnamefont {Soler-Delgado}}, \bibinfo {author} {\bibfnamefont {F.}~\bibnamefont {Yao}}, \bibinfo {author} {\bibfnamefont {D.}~\bibnamefont {Dumcenco}}, \bibinfo {author} {\bibfnamefont {E.}~\bibnamefont {Giannini}}, \bibinfo {author} {\bibfnamefont {J.}~\bibnamefont {Li}}, \bibinfo {author} {\bibfnamefont {C.~A.}\ \bibnamefont {Occhialini}}, \bibinfo {author} {\bibfnamefont {R.}~\bibnamefont {Comin}}, \bibinfo {author} {\bibfnamefont {N.}~\bibnamefont {Ubrig}},\ and\ \bibinfo {author} {\bibfnamefont {A.~F.}\ \bibnamefont {Morpurgo}},\ }\bibfield  {title} {\bibinfo {title} {Probing {Magnetism} in {Exfoliated} {VI} $_{\textrm{3}}$ {Layers} with {Magnetotransport}},\ }\href {https://doi.org/10.1021/acs.nanolett.2c01361} {\bibfield  {journal} {\bibinfo  {journal} {Nano Letters}\ }\textbf {\bibinfo {volume} {22}},\ \bibinfo {pages} {6149} (\bibinfo {year} {2022})}\BibitemShut {NoStop}%
\bibitem [{\citenamefont {Fowler}\ and\ \citenamefont {Nordheim}(1928)}]{fowler1928electron}%
  \BibitemOpen
  \bibfield  {author} {\bibinfo {author} {\bibfnamefont {R.~H.}\ \bibnamefont {Fowler}}\ and\ \bibinfo {author} {\bibfnamefont {L.}~\bibnamefont {Nordheim}},\ }\bibfield  {title} {\bibinfo {title} {Electron emission in intense electric fields},\ }\href@noop {} {\bibfield  {journal} {\bibinfo  {journal} {Proceedings of the Royal Society of London. Series A, Containing Papers of a Mathematical and Physical Character}\ }\textbf {\bibinfo {volume} {119}},\ \bibinfo {pages} {173} (\bibinfo {year} {1928})}\BibitemShut {NoStop}%
\bibitem [{\citenamefont {Lenzlinger}\ and\ \citenamefont {Snow}(1969)}]{lenzlinger1969fowler}%
  \BibitemOpen
  \bibfield  {author} {\bibinfo {author} {\bibfnamefont {M.}~\bibnamefont {Lenzlinger}}\ and\ \bibinfo {author} {\bibfnamefont {E.}~\bibnamefont {Snow}},\ }\bibfield  {title} {\bibinfo {title} {Fowler-nordheim tunneling into thermally grown sio2},\ }\href {https://doi.org/10.1063/1.1657043} {\bibfield  {journal} {\bibinfo  {journal} {Journal of Applied physics}\ }\textbf {\bibinfo {volume} {40}},\ \bibinfo {pages} {278} (\bibinfo {year} {1969})}\BibitemShut {NoStop}%
\bibitem [{\citenamefont {Wang}\ \emph {et~al.}(2011)\citenamefont {Wang}, \citenamefont {Eyert},\ and\ \citenamefont {Schwingenschlögl}}]{wang_electronic_2011}%
  \BibitemOpen
  \bibfield  {author} {\bibinfo {author} {\bibfnamefont {H.}~\bibnamefont {Wang}}, \bibinfo {author} {\bibfnamefont {V.}~\bibnamefont {Eyert}},\ and\ \bibinfo {author} {\bibfnamefont {U.}~\bibnamefont {Schwingenschlögl}},\ }\bibfield  {title} {\bibinfo {title} {Electronic structure and magnetic ordering of the semiconducting chromium trihalides {CrCl} $_{\textrm{3}}$ , {CrBr} $_{\textrm{3}}$ , and {CrI} $_{\textrm{3}}$},\ }\href {https://doi.org/10.1088/0953-8984/23/11/116003} {\bibfield  {journal} {\bibinfo  {journal} {Journal of Physics: Condensed Matter}\ }\textbf {\bibinfo {volume} {23}},\ \bibinfo {pages} {116003} (\bibinfo {year} {2011})}\BibitemShut {NoStop}%
\bibitem [{\citenamefont {Zheng}\ \emph {et~al.}(2019)\citenamefont {Zheng}, \citenamefont {Jiang}, \citenamefont {Xue}, \citenamefont {Dai},\ and\ \citenamefont {Feng}}]{zheng_ab_2019}%
  \BibitemOpen
  \bibfield  {author} {\bibinfo {author} {\bibfnamefont {Y.}~\bibnamefont {Zheng}}, \bibinfo {author} {\bibfnamefont {X.-x.}\ \bibnamefont {Jiang}}, \bibinfo {author} {\bibfnamefont {X.-x.}\ \bibnamefont {Xue}}, \bibinfo {author} {\bibfnamefont {J.}~\bibnamefont {Dai}},\ and\ \bibinfo {author} {\bibfnamefont {Y.}~\bibnamefont {Feng}},\ }\bibfield  {title} {\bibinfo {title} {Ab initio study of pressure-driven phase transition in {FePS} 3 and {FePSe} 3},\ }\href {https://doi.org/10.1103/PhysRevB.100.174102} {\bibfield  {journal} {\bibinfo  {journal} {Physical Review B}\ }\textbf {\bibinfo {volume} {100}},\ \bibinfo {pages} {174102} (\bibinfo {year} {2019})}\BibitemShut {NoStop}%
\bibitem [{\citenamefont {Wilson}\ \emph {et~al.}(2021)\citenamefont {Wilson}, \citenamefont {Lee}, \citenamefont {Cenker}, \citenamefont {Xie}, \citenamefont {Dismukes}, \citenamefont {Telford}, \citenamefont {Fonseca}, \citenamefont {Sivakumar}, \citenamefont {Dean}, \citenamefont {Cao}, \citenamefont {Roy}, \citenamefont {Xu},\ and\ \citenamefont {Zhu}}]{wilson_interlayer_2021}%
  \BibitemOpen
  \bibfield  {author} {\bibinfo {author} {\bibfnamefont {N.~P.}\ \bibnamefont {Wilson}}, \bibinfo {author} {\bibfnamefont {K.}~\bibnamefont {Lee}}, \bibinfo {author} {\bibfnamefont {J.}~\bibnamefont {Cenker}}, \bibinfo {author} {\bibfnamefont {K.}~\bibnamefont {Xie}}, \bibinfo {author} {\bibfnamefont {A.~H.}\ \bibnamefont {Dismukes}}, \bibinfo {author} {\bibfnamefont {E.~J.}\ \bibnamefont {Telford}}, \bibinfo {author} {\bibfnamefont {J.}~\bibnamefont {Fonseca}}, \bibinfo {author} {\bibfnamefont {S.}~\bibnamefont {Sivakumar}}, \bibinfo {author} {\bibfnamefont {C.}~\bibnamefont {Dean}}, \bibinfo {author} {\bibfnamefont {T.}~\bibnamefont {Cao}}, \bibinfo {author} {\bibfnamefont {X.}~\bibnamefont {Roy}}, \bibinfo {author} {\bibfnamefont {X.}~\bibnamefont {Xu}},\ and\ \bibinfo {author} {\bibfnamefont {X.}~\bibnamefont {Zhu}},\ }\bibfield  {title} {\bibinfo {title} {Interlayer electronic coupling on demand in a {2D} magnetic semiconductor},\ }\href {https://doi.org/10.1038/s41563-021-01070-8} {\bibfield  {journal}
  {\bibinfo  {journal} {Nature Materials}\ }\textbf {\bibinfo {volume} {20}},\ \bibinfo {pages} {1657} (\bibinfo {year} {2021})}\BibitemShut {NoStop}%
\bibitem [{\citenamefont {Wu}\ \emph {et~al.}(2022)\citenamefont {Wu}, \citenamefont {Gutiérrez-Lezama}, \citenamefont {López-Paz}, \citenamefont {Gibertini}, \citenamefont {Watanabe}, \citenamefont {Taniguchi}, \citenamefont {von Rohr}, \citenamefont {Ubrig},\ and\ \citenamefont {Morpurgo}}]{wu_quasi-1d_2022}%
  \BibitemOpen
  \bibfield  {author} {\bibinfo {author} {\bibfnamefont {F.}~\bibnamefont {Wu}}, \bibinfo {author} {\bibfnamefont {I.}~\bibnamefont {Gutiérrez-Lezama}}, \bibinfo {author} {\bibfnamefont {S.~A.}\ \bibnamefont {López-Paz}}, \bibinfo {author} {\bibfnamefont {M.}~\bibnamefont {Gibertini}}, \bibinfo {author} {\bibfnamefont {K.}~\bibnamefont {Watanabe}}, \bibinfo {author} {\bibfnamefont {T.}~\bibnamefont {Taniguchi}}, \bibinfo {author} {\bibfnamefont {F.~O.}\ \bibnamefont {von Rohr}}, \bibinfo {author} {\bibfnamefont {N.}~\bibnamefont {Ubrig}},\ and\ \bibinfo {author} {\bibfnamefont {A.~F.}\ \bibnamefont {Morpurgo}},\ }\bibfield  {title} {\bibinfo {title} {Quasi-{1D} {Electronic} {Transport} in a {2D} {Magnetic} {Semiconductor}},\ }\href {https://doi.org/10.1002/adma.202109759} {\bibfield  {journal} {\bibinfo  {journal} {Advanced Materials}\ }\textbf {\bibinfo {volume} {34}},\ \bibinfo {pages} {2109759} (\bibinfo {year} {2022})}\BibitemShut {NoStop}%
\bibitem [{\citenamefont {Telford}\ \emph {et~al.}(2020)\citenamefont {Telford}, \citenamefont {Dismukes}, \citenamefont {Lee}, \citenamefont {Cheng}, \citenamefont {Wieteska}, \citenamefont {Bartholomew}, \citenamefont {Chen}, \citenamefont {Xu}, \citenamefont {Pasupathy}, \citenamefont {Zhu}, \citenamefont {Dean},\ and\ \citenamefont {Roy}}]{telford_layered_2020}%
  \BibitemOpen
  \bibfield  {author} {\bibinfo {author} {\bibfnamefont {E.~J.}\ \bibnamefont {Telford}}, \bibinfo {author} {\bibfnamefont {A.~H.}\ \bibnamefont {Dismukes}}, \bibinfo {author} {\bibfnamefont {K.}~\bibnamefont {Lee}}, \bibinfo {author} {\bibfnamefont {M.}~\bibnamefont {Cheng}}, \bibinfo {author} {\bibfnamefont {A.}~\bibnamefont {Wieteska}}, \bibinfo {author} {\bibfnamefont {A.~K.}\ \bibnamefont {Bartholomew}}, \bibinfo {author} {\bibfnamefont {Y.}~\bibnamefont {Chen}}, \bibinfo {author} {\bibfnamefont {X.}~\bibnamefont {Xu}}, \bibinfo {author} {\bibfnamefont {A.~N.}\ \bibnamefont {Pasupathy}}, \bibinfo {author} {\bibfnamefont {X.}~\bibnamefont {Zhu}}, \bibinfo {author} {\bibfnamefont {C.~R.}\ \bibnamefont {Dean}},\ and\ \bibinfo {author} {\bibfnamefont {X.}~\bibnamefont {Roy}},\ }\bibfield  {title} {\bibinfo {title} {Layered {Antiferromagnetism} {Induces} {Large} {Negative} {Magnetoresistance} in the van der {Waals} {Semiconductor} {CrSBr}},\ }\href {https://doi.org/10.1002/adma.202003240} {\bibfield
  {journal} {\bibinfo  {journal} {Advanced Materials}\ }\textbf {\bibinfo {volume} {32}},\ \bibinfo {pages} {2003240} (\bibinfo {year} {2020})}\BibitemShut {NoStop}%
\bibitem [{\citenamefont {Boix-Constant}\ \emph {et~al.}(2022)\citenamefont {Boix-Constant}, \citenamefont {Mañas-Valero}, \citenamefont {Ruiz}, \citenamefont {Rybakov}, \citenamefont {Konieczny}, \citenamefont {Pillet}, \citenamefont {Baldoví},\ and\ \citenamefont {Coronado}}]{boix-constant_probing_2022}%
  \BibitemOpen
  \bibfield  {author} {\bibinfo {author} {\bibfnamefont {C.}~\bibnamefont {Boix-Constant}}, \bibinfo {author} {\bibfnamefont {S.}~\bibnamefont {Mañas-Valero}}, \bibinfo {author} {\bibfnamefont {A.~M.}\ \bibnamefont {Ruiz}}, \bibinfo {author} {\bibfnamefont {A.}~\bibnamefont {Rybakov}}, \bibinfo {author} {\bibfnamefont {K.~A.}\ \bibnamefont {Konieczny}}, \bibinfo {author} {\bibfnamefont {S.}~\bibnamefont {Pillet}}, \bibinfo {author} {\bibfnamefont {J.~J.}\ \bibnamefont {Baldoví}},\ and\ \bibinfo {author} {\bibfnamefont {E.}~\bibnamefont {Coronado}},\ }\bibfield  {title} {\bibinfo {title} {Probing the spin dimensionality in single-layer crsbr van der waals heterostructures by magneto-transport measurements},\ }\href {https://doi.org/https://doi.org/10.1002/adma.202204940} {\bibfield  {journal} {\bibinfo  {journal} {Advanced Materials}\ }\textbf {\bibinfo {volume} {34}},\ \bibinfo {pages} {2204940} (\bibinfo {year} {2022})}\BibitemShut {NoStop}%
\bibitem [{\citenamefont {Telford}\ \emph {et~al.}(2022)\citenamefont {Telford}, \citenamefont {Dismukes}, \citenamefont {Dudley}, \citenamefont {Wiscons}, \citenamefont {Lee}, \citenamefont {Chica}, \citenamefont {Ziebel}, \citenamefont {Han}, \citenamefont {Yu}, \citenamefont {Shabani}, \citenamefont {Scheie}, \citenamefont {Watanabe}, \citenamefont {Taniguchi}, \citenamefont {Xiao}, \citenamefont {Zhu}, \citenamefont {Pasupathy}, \citenamefont {Nuckolls}, \citenamefont {Zhu}, \citenamefont {Dean},\ and\ \citenamefont {Roy}}]{telford_coupling_2022}%
  \BibitemOpen
  \bibfield  {author} {\bibinfo {author} {\bibfnamefont {E.~J.}\ \bibnamefont {Telford}}, \bibinfo {author} {\bibfnamefont {A.~H.}\ \bibnamefont {Dismukes}}, \bibinfo {author} {\bibfnamefont {R.~L.}\ \bibnamefont {Dudley}}, \bibinfo {author} {\bibfnamefont {R.~A.}\ \bibnamefont {Wiscons}}, \bibinfo {author} {\bibfnamefont {K.}~\bibnamefont {Lee}}, \bibinfo {author} {\bibfnamefont {D.~G.}\ \bibnamefont {Chica}}, \bibinfo {author} {\bibfnamefont {M.~E.}\ \bibnamefont {Ziebel}}, \bibinfo {author} {\bibfnamefont {M.-G.}\ \bibnamefont {Han}}, \bibinfo {author} {\bibfnamefont {J.}~\bibnamefont {Yu}}, \bibinfo {author} {\bibfnamefont {S.}~\bibnamefont {Shabani}}, \bibinfo {author} {\bibfnamefont {A.}~\bibnamefont {Scheie}}, \bibinfo {author} {\bibfnamefont {K.}~\bibnamefont {Watanabe}}, \bibinfo {author} {\bibfnamefont {T.}~\bibnamefont {Taniguchi}}, \bibinfo {author} {\bibfnamefont {D.}~\bibnamefont {Xiao}}, \bibinfo {author} {\bibfnamefont {Y.}~\bibnamefont {Zhu}}, \bibinfo {author} {\bibfnamefont {A.~N.}\
  \bibnamefont {Pasupathy}}, \bibinfo {author} {\bibfnamefont {C.}~\bibnamefont {Nuckolls}}, \bibinfo {author} {\bibfnamefont {X.}~\bibnamefont {Zhu}}, \bibinfo {author} {\bibfnamefont {C.~R.}\ \bibnamefont {Dean}},\ and\ \bibinfo {author} {\bibfnamefont {X.}~\bibnamefont {Roy}},\ }\bibfield  {title} {\bibinfo {title} {Coupling between magnetic order and charge transport in a two-dimensional magnetic semiconductor},\ }\href {https://doi.org/10.1038/s41563-022-01245-x} {\bibfield  {journal} {\bibinfo  {journal} {Nature Materials}\ }\textbf {\bibinfo {volume} {21}},\ \bibinfo {pages} {754} (\bibinfo {year} {2022})}\BibitemShut {NoStop}%
\bibitem [{\citenamefont {Klein}\ \emph {et~al.}(2023)\citenamefont {Klein}, \citenamefont {Song}, \citenamefont {Pingault}, \citenamefont {Dirnberger}, \citenamefont {Chi}, \citenamefont {Curtis}, \citenamefont {Dana}, \citenamefont {Bushati}, \citenamefont {Quan}, \citenamefont {Dekanovsky}, \citenamefont {Sofer}, \citenamefont {Alù}, \citenamefont {Menon}, \citenamefont {Moodera}, \citenamefont {Lončar}, \citenamefont {Narang},\ and\ \citenamefont {Ross}}]{Kleinacsnano_2023}%
  \BibitemOpen
  \bibfield  {author} {\bibinfo {author} {\bibfnamefont {J.}~\bibnamefont {Klein}}, \bibinfo {author} {\bibfnamefont {Z.}~\bibnamefont {Song}}, \bibinfo {author} {\bibfnamefont {B.}~\bibnamefont {Pingault}}, \bibinfo {author} {\bibfnamefont {F.}~\bibnamefont {Dirnberger}}, \bibinfo {author} {\bibfnamefont {H.}~\bibnamefont {Chi}}, \bibinfo {author} {\bibfnamefont {J.~B.}\ \bibnamefont {Curtis}}, \bibinfo {author} {\bibfnamefont {R.}~\bibnamefont {Dana}}, \bibinfo {author} {\bibfnamefont {R.}~\bibnamefont {Bushati}}, \bibinfo {author} {\bibfnamefont {J.}~\bibnamefont {Quan}}, \bibinfo {author} {\bibfnamefont {L.}~\bibnamefont {Dekanovsky}}, \bibinfo {author} {\bibfnamefont {Z.}~\bibnamefont {Sofer}}, \bibinfo {author} {\bibfnamefont {A.}~\bibnamefont {Alù}}, \bibinfo {author} {\bibfnamefont {V.~M.}\ \bibnamefont {Menon}}, \bibinfo {author} {\bibfnamefont {J.~S.}\ \bibnamefont {Moodera}}, \bibinfo {author} {\bibfnamefont {M.}~\bibnamefont {Lončar}}, \bibinfo {author} {\bibfnamefont {P.}~\bibnamefont
  {Narang}},\ and\ \bibinfo {author} {\bibfnamefont {F.~M.}\ \bibnamefont {Ross}},\ }\bibfield  {title} {\bibinfo {title} {Sensing the local magnetic environment through optically active defects in a layered magnetic semiconductor},\ }\href {https://doi.org/10.1021/acsnano.2c07655} {\bibfield  {journal} {\bibinfo  {journal} {ACS Nano}\ }\textbf {\bibinfo {volume} {17}},\ \bibinfo {pages} {288} (\bibinfo {year} {2023})}\BibitemShut {NoStop}%
\bibitem [{\citenamefont {López-Paz}\ \emph {et~al.}(2022)\citenamefont {López-Paz}, \citenamefont {Guguchia}, \citenamefont {Pomjakushin}, \citenamefont {Witteveen}, \citenamefont {Cervellino}, \citenamefont {Luetkens}, \citenamefont {Casati}, \citenamefont {Morpurgo},\ and\ \citenamefont {von Rohr}}]{Sara_2022}%
  \BibitemOpen
  \bibfield  {author} {\bibinfo {author} {\bibfnamefont {S.~A.}\ \bibnamefont {López-Paz}}, \bibinfo {author} {\bibfnamefont {Z.}~\bibnamefont {Guguchia}}, \bibinfo {author} {\bibfnamefont {V.~Y.}\ \bibnamefont {Pomjakushin}}, \bibinfo {author} {\bibfnamefont {C.}~\bibnamefont {Witteveen}}, \bibinfo {author} {\bibfnamefont {A.}~\bibnamefont {Cervellino}}, \bibinfo {author} {\bibfnamefont {H.}~\bibnamefont {Luetkens}}, \bibinfo {author} {\bibfnamefont {N.}~\bibnamefont {Casati}}, \bibinfo {author} {\bibfnamefont {A.~F.}\ \bibnamefont {Morpurgo}},\ and\ \bibinfo {author} {\bibfnamefont {F.~O.}\ \bibnamefont {von Rohr}},\ }\bibfield  {title} {\bibinfo {title} {Dynamic magnetic crossover at the origin of the hidden-order in van der waals antiferromagnet crsbr},\ }\bibfield  {journal} {\bibinfo  {journal} {Nature Communications}\ }\textbf {\bibinfo {volume} {13}},\ \href {https://doi.org/10.1038/s41467-022-32290-4} {10.1038/s41467-022-32290-4} (\bibinfo {year} {2022})\BibitemShut {NoStop}%
\bibitem [{\citenamefont {Deng}\ \emph {et~al.}(2021)\citenamefont {Deng}, \citenamefont {Guo}, \citenamefont {Hosono}, \citenamefont {Ying},\ and\ \citenamefont {Chen}}]{deng_two-dimensional_2021}%
  \BibitemOpen
  \bibfield  {author} {\bibinfo {author} {\bibfnamefont {J.}~\bibnamefont {Deng}}, \bibinfo {author} {\bibfnamefont {J.}~\bibnamefont {Guo}}, \bibinfo {author} {\bibfnamefont {H.}~\bibnamefont {Hosono}}, \bibinfo {author} {\bibfnamefont {T.}~\bibnamefont {Ying}},\ and\ \bibinfo {author} {\bibfnamefont {X.}~\bibnamefont {Chen}},\ }\bibfield  {title} {\bibinfo {title} {Two-dimensional bipolar ferromagnetic semiconductors from layered antiferromagnets},\ }\href {https://doi.org/10.1103/PhysRevMaterials.5.034005} {\bibfield  {journal} {\bibinfo  {journal} {Physical Review Materials}\ }\textbf {\bibinfo {volume} {5}},\ \bibinfo {pages} {034005} (\bibinfo {year} {2021})}\BibitemShut {NoStop}%
\bibitem [{\citenamefont {Najmaei}\ \emph {et~al.}(2018)\citenamefont {Najmaei}, \citenamefont {Neupane}, \citenamefont {Nichols}, \citenamefont {Burke}, \citenamefont {Mazzoni}, \citenamefont {Chin}, \citenamefont {Rhodes}, \citenamefont {Balicas}, \citenamefont {Franklin},\ and\ \citenamefont {Dubey}}]{najmaei_cross-plane_2018}%
  \BibitemOpen
  \bibfield  {author} {\bibinfo {author} {\bibfnamefont {S.}~\bibnamefont {Najmaei}}, \bibinfo {author} {\bibfnamefont {M.~R.}\ \bibnamefont {Neupane}}, \bibinfo {author} {\bibfnamefont {B.~M.}\ \bibnamefont {Nichols}}, \bibinfo {author} {\bibfnamefont {R.~A.}\ \bibnamefont {Burke}}, \bibinfo {author} {\bibfnamefont {A.~L.}\ \bibnamefont {Mazzoni}}, \bibinfo {author} {\bibfnamefont {M.~L.}\ \bibnamefont {Chin}}, \bibinfo {author} {\bibfnamefont {D.~A.}\ \bibnamefont {Rhodes}}, \bibinfo {author} {\bibfnamefont {L.}~\bibnamefont {Balicas}}, \bibinfo {author} {\bibfnamefont {A.~D.}\ \bibnamefont {Franklin}},\ and\ \bibinfo {author} {\bibfnamefont {M.}~\bibnamefont {Dubey}},\ }\bibfield  {title} {\bibinfo {title} {Cross-{Plane} {Carrier} {Transport} in {Van} der {Waals} {Layered} {Materials}},\ }\href {https://doi.org/10.1002/smll.201703808} {\bibfield  {journal} {\bibinfo  {journal} {Small}\ }\textbf {\bibinfo {volume} {14}},\ \bibinfo {pages} {1703808} (\bibinfo {year} {2018})}\BibitemShut {NoStop}%
\bibitem [{\citenamefont {Shklovskii}\ and\ \citenamefont {Efros}(2013)}]{shklovskii2013electronic}%
  \BibitemOpen
  \bibfield  {author} {\bibinfo {author} {\bibfnamefont {B.~I.}\ \bibnamefont {Shklovskii}}\ and\ \bibinfo {author} {\bibfnamefont {A.~L.}\ \bibnamefont {Efros}},\ }\href@noop {} {\emph {\bibinfo {title} {Electronic properties of doped semiconductors}}},\ Vol.~\bibinfo {volume} {45}\ (\bibinfo  {publisher} {Springer Science \& Business Media},\ \bibinfo {year} {2013})\BibitemShut {NoStop}%
\bibitem [{\citenamefont {Pollak}\ and\ \citenamefont {Shklovskii}(1991)}]{pollak1991hopping}%
  \BibitemOpen
  \bibfield  {author} {\bibinfo {author} {\bibfnamefont {M.}~\bibnamefont {Pollak}}\ and\ \bibinfo {author} {\bibfnamefont {B.}~\bibnamefont {Shklovskii}},\ }\href@noop {} {\emph {\bibinfo {title} {Hopping transport in solids}}}\ (\bibinfo  {publisher} {Elsevier},\ \bibinfo {year} {1991})\BibitemShut {NoStop}%
\bibitem [{\citenamefont {Wu}\ \emph {et~al.}(2023{\natexlab{a}})\citenamefont {Wu}, \citenamefont {Gibertini}, \citenamefont {Watanabe}, \citenamefont {Taniguchi}, \citenamefont {Gutiérrez-Lezama}, \citenamefont {Ubrig},\ and\ \citenamefont {Morpurgo}}]{wu_gate-controlled_2023}%
  \BibitemOpen
  \bibfield  {author} {\bibinfo {author} {\bibfnamefont {F.}~\bibnamefont {Wu}}, \bibinfo {author} {\bibfnamefont {M.}~\bibnamefont {Gibertini}}, \bibinfo {author} {\bibfnamefont {K.}~\bibnamefont {Watanabe}}, \bibinfo {author} {\bibfnamefont {T.}~\bibnamefont {Taniguchi}}, \bibinfo {author} {\bibfnamefont {I.}~\bibnamefont {Gutiérrez-Lezama}}, \bibinfo {author} {\bibfnamefont {N.}~\bibnamefont {Ubrig}},\ and\ \bibinfo {author} {\bibfnamefont {A.~F.}\ \bibnamefont {Morpurgo}},\ }\bibfield  {title} {\bibinfo {title} {Gate-{Controlled} {Magnetotransport} and {Electrostatic} {Modulation} of {Magnetism} in {2D} {Magnetic} {Semiconductor} {CrPS4}},\ }\href {https://doi.org/10.1002/adma.202211653} {\bibfield  {journal} {\bibinfo  {journal} {Advanced Materials}\ }\textbf {\bibinfo {volume} {35}},\ \bibinfo {pages} {2211653} (\bibinfo {year} {2023}{\natexlab{a}})}\BibitemShut {NoStop}%
\bibitem [{\citenamefont {Wu}\ \emph {et~al.}(2023{\natexlab{b}})\citenamefont {Wu}, \citenamefont {Gibertini}, \citenamefont {Watanabe}, \citenamefont {Taniguchi}, \citenamefont {Gutiérrez-Lezama}, \citenamefont {Ubrig},\ and\ \citenamefont {Morpurgo}}]{wu_magnetism-induced_2023}%
  \BibitemOpen
  \bibfield  {author} {\bibinfo {author} {\bibfnamefont {F.}~\bibnamefont {Wu}}, \bibinfo {author} {\bibfnamefont {M.}~\bibnamefont {Gibertini}}, \bibinfo {author} {\bibfnamefont {K.}~\bibnamefont {Watanabe}}, \bibinfo {author} {\bibfnamefont {T.}~\bibnamefont {Taniguchi}}, \bibinfo {author} {\bibfnamefont {I.}~\bibnamefont {Gutiérrez-Lezama}}, \bibinfo {author} {\bibfnamefont {N.}~\bibnamefont {Ubrig}},\ and\ \bibinfo {author} {\bibfnamefont {A.~F.}\ \bibnamefont {Morpurgo}},\ }\bibfield  {title} {\bibinfo {title} {Magnetism-induced band-edge shift as mechanism for magnetoconductance in {CrPS}4 transistors},\ }\bibfield  {journal} {\bibinfo  {journal} {Nano Letters}\ }\href {https://doi.org/10.1021/acs.nanolett.3c02274} {10.1021/acs.nanolett.3c02274} (\bibinfo {year} {2023}{\natexlab{b}})\BibitemShut {NoStop}%
\bibitem [{\citenamefont {Giannozzi}\ \emph {et~al.}(2009)\citenamefont {Giannozzi}, \citenamefont {Baroni}, \citenamefont {Bonini}, \citenamefont {Calandra}, \citenamefont {Car}, \citenamefont {Cavazzoni}, \citenamefont {{Davide Ceresoli}}, \citenamefont {Chiarotti}, \citenamefont {Cococcioni}, \citenamefont {Dabo}, \citenamefont {Corso}, \citenamefont {Gironcoli}, \citenamefont {Fabris}, \citenamefont {Fratesi}, \citenamefont {Gebauer}, \citenamefont {Gerstmann}, \citenamefont {Gougoussis}, \citenamefont {{Anton Kokalj}}, \citenamefont {Lazzeri}, \citenamefont {Martin-Samos}, \citenamefont {Marzari}, \citenamefont {Mauri}, \citenamefont {Mazzarello}, \citenamefont {{Stefano Paolini}}, \citenamefont {Pasquarello}, \citenamefont {Paulatto}, \citenamefont {Sbraccia}, \citenamefont {Scandolo}, \citenamefont {Sclauzero}, \citenamefont {Seitsonen}, \citenamefont {Smogunov}, \citenamefont {Umari},\ and\ \citenamefont {Wentzcovitch}}]{giannozzi_quantum_2009}%
  \BibitemOpen
  \bibfield  {author} {\bibinfo {author} {\bibfnamefont {P.}~\bibnamefont {Giannozzi}}, \bibinfo {author} {\bibfnamefont {S.}~\bibnamefont {Baroni}}, \bibinfo {author} {\bibfnamefont {N.}~\bibnamefont {Bonini}}, \bibinfo {author} {\bibfnamefont {M.}~\bibnamefont {Calandra}}, \bibinfo {author} {\bibfnamefont {R.}~\bibnamefont {Car}}, \bibinfo {author} {\bibfnamefont {C.}~\bibnamefont {Cavazzoni}}, \bibinfo {author} {\bibnamefont {{Davide Ceresoli}}}, \bibinfo {author} {\bibfnamefont {G.~L.}\ \bibnamefont {Chiarotti}}, \bibinfo {author} {\bibfnamefont {M.}~\bibnamefont {Cococcioni}}, \bibinfo {author} {\bibfnamefont {I.}~\bibnamefont {Dabo}}, \bibinfo {author} {\bibfnamefont {A.~D.}\ \bibnamefont {Corso}}, \bibinfo {author} {\bibfnamefont {S.~d.}\ \bibnamefont {Gironcoli}}, \bibinfo {author} {\bibfnamefont {S.}~\bibnamefont {Fabris}}, \bibinfo {author} {\bibfnamefont {G.}~\bibnamefont {Fratesi}}, \bibinfo {author} {\bibfnamefont {R.}~\bibnamefont {Gebauer}}, \bibinfo {author} {\bibfnamefont {U.}~\bibnamefont
  {Gerstmann}}, \bibinfo {author} {\bibfnamefont {C.}~\bibnamefont {Gougoussis}}, \bibinfo {author} {\bibnamefont {{Anton Kokalj}}}, \bibinfo {author} {\bibfnamefont {M.}~\bibnamefont {Lazzeri}}, \bibinfo {author} {\bibfnamefont {L.}~\bibnamefont {Martin-Samos}}, \bibinfo {author} {\bibfnamefont {N.}~\bibnamefont {Marzari}}, \bibinfo {author} {\bibfnamefont {F.}~\bibnamefont {Mauri}}, \bibinfo {author} {\bibfnamefont {R.}~\bibnamefont {Mazzarello}}, \bibinfo {author} {\bibnamefont {{Stefano Paolini}}}, \bibinfo {author} {\bibfnamefont {A.}~\bibnamefont {Pasquarello}}, \bibinfo {author} {\bibfnamefont {L.}~\bibnamefont {Paulatto}}, \bibinfo {author} {\bibfnamefont {C.}~\bibnamefont {Sbraccia}}, \bibinfo {author} {\bibfnamefont {S.}~\bibnamefont {Scandolo}}, \bibinfo {author} {\bibfnamefont {G.}~\bibnamefont {Sclauzero}}, \bibinfo {author} {\bibfnamefont {A.~P.}\ \bibnamefont {Seitsonen}}, \bibinfo {author} {\bibfnamefont {A.}~\bibnamefont {Smogunov}}, \bibinfo {author} {\bibfnamefont {P.}~\bibnamefont
  {Umari}},\ and\ \bibinfo {author} {\bibfnamefont {R.~M.}\ \bibnamefont {Wentzcovitch}},\ }\bibfield  {title} {\bibinfo {title} {{QUANTUM} {ESPRESSO}: a modular and open-source software project for quantum simulations of materials},\ }\href {https://doi.org/10.1088/0953-8984/21/39/395502} {\bibfield  {journal} {\bibinfo  {journal} {Journal of Physics: Condensed Matter}\ }\textbf {\bibinfo {volume} {21}},\ \bibinfo {pages} {395502} (\bibinfo {year} {2009})}\BibitemShut {NoStop}%
\bibitem [{\citenamefont {Giannozzi}\ \emph {et~al.}(2017)\citenamefont {Giannozzi}, \citenamefont {Andreussi}, \citenamefont {Brumme}, \citenamefont {Bunau}, \citenamefont {Buongiorno~Nardelli}, \citenamefont {Calandra}, \citenamefont {Car}, \citenamefont {Cavazzoni}, \citenamefont {Ceresoli}, \citenamefont {Cococcioni}, \citenamefont {Colonna}, \citenamefont {Carnimeo}, \citenamefont {Dal~Corso}, \citenamefont {De~Gironcoli}, \citenamefont {Delugas}, \citenamefont {Distasio}, \citenamefont {Ferretti}, \citenamefont {Floris}, \citenamefont {Fratesi}, \citenamefont {Fugallo}, \citenamefont {Gebauer}, \citenamefont {Gerstmann}, \citenamefont {Giustino}, \citenamefont {Gorni}, \citenamefont {Jia}, \citenamefont {Kawamura}, \citenamefont {Ko}, \citenamefont {Kokalj}, \citenamefont {Kücükbenli}, \citenamefont {Lazzeri}, \citenamefont {Marsili}, \citenamefont {Marzari}, \citenamefont {Mauri}, \citenamefont {Nguyen}, \citenamefont {Nguyen}, \citenamefont {Otero-De-La-Roza}, \citenamefont {Paulatto}, \citenamefont
  {Poncé}, \citenamefont {Rocca}, \citenamefont {Sabatini}, \citenamefont {Santra}, \citenamefont {Schlipf}, \citenamefont {Seitsonen}, \citenamefont {Smogunov}, \citenamefont {Timrov}, \citenamefont {Thonhauser}, \citenamefont {Umari}, \citenamefont {Vast}, \citenamefont {Wu},\ and\ \citenamefont {Baroni}}]{giannozzi_advanced_2017}%
  \BibitemOpen
  \bibfield  {author} {\bibinfo {author} {\bibfnamefont {P.}~\bibnamefont {Giannozzi}}, \bibinfo {author} {\bibfnamefont {O.}~\bibnamefont {Andreussi}}, \bibinfo {author} {\bibfnamefont {T.}~\bibnamefont {Brumme}}, \bibinfo {author} {\bibfnamefont {O.}~\bibnamefont {Bunau}}, \bibinfo {author} {\bibfnamefont {M.}~\bibnamefont {Buongiorno~Nardelli}}, \bibinfo {author} {\bibfnamefont {M.}~\bibnamefont {Calandra}}, \bibinfo {author} {\bibfnamefont {R.}~\bibnamefont {Car}}, \bibinfo {author} {\bibfnamefont {C.}~\bibnamefont {Cavazzoni}}, \bibinfo {author} {\bibfnamefont {D.}~\bibnamefont {Ceresoli}}, \bibinfo {author} {\bibfnamefont {M.}~\bibnamefont {Cococcioni}}, \bibinfo {author} {\bibfnamefont {N.}~\bibnamefont {Colonna}}, \bibinfo {author} {\bibfnamefont {I.}~\bibnamefont {Carnimeo}}, \bibinfo {author} {\bibfnamefont {A.}~\bibnamefont {Dal~Corso}}, \bibinfo {author} {\bibfnamefont {S.}~\bibnamefont {De~Gironcoli}}, \bibinfo {author} {\bibfnamefont {P.}~\bibnamefont {Delugas}}, \bibinfo {author} {\bibfnamefont
  {R.~A.}\ \bibnamefont {Distasio}}, \bibinfo {author} {\bibfnamefont {A.}~\bibnamefont {Ferretti}}, \bibinfo {author} {\bibfnamefont {A.}~\bibnamefont {Floris}}, \bibinfo {author} {\bibfnamefont {G.}~\bibnamefont {Fratesi}}, \bibinfo {author} {\bibfnamefont {G.}~\bibnamefont {Fugallo}}, \bibinfo {author} {\bibfnamefont {R.}~\bibnamefont {Gebauer}}, \bibinfo {author} {\bibfnamefont {U.}~\bibnamefont {Gerstmann}}, \bibinfo {author} {\bibfnamefont {F.}~\bibnamefont {Giustino}}, \bibinfo {author} {\bibfnamefont {T.}~\bibnamefont {Gorni}}, \bibinfo {author} {\bibfnamefont {J.}~\bibnamefont {Jia}}, \bibinfo {author} {\bibfnamefont {M.}~\bibnamefont {Kawamura}}, \bibinfo {author} {\bibfnamefont {H.~Y.}\ \bibnamefont {Ko}}, \bibinfo {author} {\bibfnamefont {A.}~\bibnamefont {Kokalj}}, \bibinfo {author} {\bibfnamefont {E.}~\bibnamefont {Kücükbenli}}, \bibinfo {author} {\bibfnamefont {M.}~\bibnamefont {Lazzeri}}, \bibinfo {author} {\bibfnamefont {M.}~\bibnamefont {Marsili}}, \bibinfo {author} {\bibfnamefont
  {N.}~\bibnamefont {Marzari}}, \bibinfo {author} {\bibfnamefont {F.}~\bibnamefont {Mauri}}, \bibinfo {author} {\bibfnamefont {N.~L.}\ \bibnamefont {Nguyen}}, \bibinfo {author} {\bibfnamefont {H.~V.}\ \bibnamefont {Nguyen}}, \bibinfo {author} {\bibfnamefont {A.}~\bibnamefont {Otero-De-La-Roza}}, \bibinfo {author} {\bibfnamefont {L.}~\bibnamefont {Paulatto}}, \bibinfo {author} {\bibfnamefont {S.}~\bibnamefont {Poncé}}, \bibinfo {author} {\bibfnamefont {D.}~\bibnamefont {Rocca}}, \bibinfo {author} {\bibfnamefont {R.}~\bibnamefont {Sabatini}}, \bibinfo {author} {\bibfnamefont {B.}~\bibnamefont {Santra}}, \bibinfo {author} {\bibfnamefont {M.}~\bibnamefont {Schlipf}}, \bibinfo {author} {\bibfnamefont {A.~P.}\ \bibnamefont {Seitsonen}}, \bibinfo {author} {\bibfnamefont {A.}~\bibnamefont {Smogunov}}, \bibinfo {author} {\bibfnamefont {I.}~\bibnamefont {Timrov}}, \bibinfo {author} {\bibfnamefont {T.}~\bibnamefont {Thonhauser}}, \bibinfo {author} {\bibfnamefont {P.}~\bibnamefont {Umari}}, \bibinfo {author}
  {\bibfnamefont {N.}~\bibnamefont {Vast}}, \bibinfo {author} {\bibfnamefont {X.}~\bibnamefont {Wu}},\ and\ \bibinfo {author} {\bibfnamefont {S.}~\bibnamefont {Baroni}},\ }\bibfield  {title} {\bibinfo {title} {Advanced {Capabilities} for {Materials} {Modelling} with {Quantum} {ESPRESSO}},\ }\href {https://doi.org/10.1088/1361-648X/aa8f79} {\bibfield  {journal} {\bibinfo  {journal} {Journal of Physics Condensed Matter}\ }\textbf {\bibinfo {volume} {29}},\ \bibinfo {pages} {465901} (\bibinfo {year} {2017})}\BibitemShut {NoStop}%
\bibitem [{\citenamefont {Thonhauser}\ \emph {et~al.}(2015)\citenamefont {Thonhauser}, \citenamefont {Zuluaga}, \citenamefont {Arter}, \citenamefont {Berland}, \citenamefont {Schröder},\ and\ \citenamefont {Hyldgaard}}]{thonhauser_spin_2015}%
  \BibitemOpen
  \bibfield  {author} {\bibinfo {author} {\bibfnamefont {T.}~\bibnamefont {Thonhauser}}, \bibinfo {author} {\bibfnamefont {S.}~\bibnamefont {Zuluaga}}, \bibinfo {author} {\bibfnamefont {C.}~\bibnamefont {Arter}}, \bibinfo {author} {\bibfnamefont {K.}~\bibnamefont {Berland}}, \bibinfo {author} {\bibfnamefont {E.}~\bibnamefont {Schröder}},\ and\ \bibinfo {author} {\bibfnamefont {P.}~\bibnamefont {Hyldgaard}},\ }\bibfield  {title} {\bibinfo {title} {Spin {Signature} of {Nonlocal} {Correlation} {Binding} in {Metal}-{Organic} {Frameworks}},\ }\href {https://doi.org/10.1103/PhysRevLett.115.136402} {\bibfield  {journal} {\bibinfo  {journal} {Physical Review Letters}\ }\textbf {\bibinfo {volume} {115}},\ \bibinfo {pages} {136402} (\bibinfo {year} {2015})}\BibitemShut {NoStop}%
\bibitem [{\citenamefont {Chakraborty}\ \emph {et~al.}(2020)\citenamefont {Chakraborty}, \citenamefont {Berland},\ and\ \citenamefont {Thonhauser}}]{chakraborty_next_2020}%
  \BibitemOpen
  \bibfield  {author} {\bibinfo {author} {\bibfnamefont {D.}~\bibnamefont {Chakraborty}}, \bibinfo {author} {\bibfnamefont {K.}~\bibnamefont {Berland}},\ and\ \bibinfo {author} {\bibfnamefont {T.}~\bibnamefont {Thonhauser}},\ }\bibfield  {title} {\bibinfo {title} {Next-generation nonlocal van der waals density functional},\ }\href {https://doi.org/10.1021/acs.jctc.0c00471} {\bibfield  {journal} {\bibinfo  {journal} {Journal of Chemical Theory and Computation}\ }\textbf {\bibinfo {volume} {16}},\ \bibinfo {pages} {5893} (\bibinfo {year} {2020})}\BibitemShut {NoStop}%
\bibitem [{\citenamefont {Prandini}\ \emph {et~al.}(2018)\citenamefont {Prandini}, \citenamefont {Marrazzo}, \citenamefont {Castelli}, \citenamefont {Mounet},\ and\ \citenamefont {Marzari}}]{prandini_precision_2018}%
  \BibitemOpen
  \bibfield  {author} {\bibinfo {author} {\bibfnamefont {G.}~\bibnamefont {Prandini}}, \bibinfo {author} {\bibfnamefont {A.}~\bibnamefont {Marrazzo}}, \bibinfo {author} {\bibfnamefont {I.~E.}\ \bibnamefont {Castelli}}, \bibinfo {author} {\bibfnamefont {N.}~\bibnamefont {Mounet}},\ and\ \bibinfo {author} {\bibfnamefont {N.}~\bibnamefont {Marzari}},\ }\bibfield  {title} {\bibinfo {title} {Precision and {Efficiency} in {Solid}-{State} {Pseudopotential} {Calculations}},\ }\href {https://doi.org/10.1038/s41524-018-0127-2} {\bibfield  {journal} {\bibinfo  {journal} {npj Computational Materials}\ }\textbf {\bibinfo {volume} {4}},\ \bibinfo {pages} {72} (\bibinfo {year} {2018})}\BibitemShut {NoStop}%
\bibitem [{\citenamefont {Garrity}\ \emph {et~al.}(2014)\citenamefont {Garrity}, \citenamefont {Bennett}, \citenamefont {Rabe},\ and\ \citenamefont {Vanderbilt}}]{garrity_pseudopotentials_2014}%
  \BibitemOpen
  \bibfield  {author} {\bibinfo {author} {\bibfnamefont {K.~F.}\ \bibnamefont {Garrity}}, \bibinfo {author} {\bibfnamefont {J.~W.}\ \bibnamefont {Bennett}}, \bibinfo {author} {\bibfnamefont {K.~M.}\ \bibnamefont {Rabe}},\ and\ \bibinfo {author} {\bibfnamefont {D.}~\bibnamefont {Vanderbilt}},\ }\bibfield  {title} {\bibinfo {title} {Pseudopotentials for high-throughput {DFT} calculations},\ }\href {https://doi.org/https://doi.org/10.1016/j.commatsci.2013.08.053} {\bibfield  {journal} {\bibinfo  {journal} {Computational Materials Science}\ }\textbf {\bibinfo {volume} {81}},\ \bibinfo {pages} {446} (\bibinfo {year} {2014})}\BibitemShut {NoStop}%
\end{thebibliography}
%

\end{document}